# Magnetism of the 2D honeycomb layered Na$_2$Ni$_2$TeO$_6$ compound driven by intermediate Na-layer crystal-structure

A. K. Bera, and S. M. Yusuf
*Solid State Physics Division, Bhabha Atomic Research Centre, Mumbai 40085, India*
*and Homi Bhabha National Institute, Anushaktinagar, Mumbai 400094, India*

L. Keller
*Laboratory for Neutron Scattering and Imaging, Paul Scherrer Institut, CH-5232 Villigen PSI, Switzerland*

F. Yokaichiya
*Helmholtz-Zentrum Berlin für Materialien und Energie, 14109 Berlin, Germany*

J. R. Stewart
*ISIS Facility, STFC Rutherford Appleton Laboratory, Harwell Oxford, Didcot OX11 0QX, United Kingdom*

**Abstract:** The microscopic spin-spin correlations in the 2D layered spin-1 honeycomb lattice compound Na$_2$Ni$_2$TeO$_6$ have been investigated by neutron diffraction and inelastic neutron scattering. The present study reveals a novel magnetic phenomenon where the magnetic symmetry is controlled by the nonmagnetic Na-layer which is a unique feature for the studied compound Na$_2$Ni$_2$TeO$_6$ with respect to other Na based layered compounds, especially $A_2M_2XO_6$ or $A_3M_2XO_6$ compounds. The honeycomb lattice of spin-1 Ni$^{2+}$ ions, within the crystallographic *ab* planes, are well separated (~ 5.6 Å) along the *c* axis by an intermediate Na layer whose crystal structure contains chiral nuclear density distributions of Na ions. The chirality of the alternating Na layers is opposite. Such alternating chirality of the Na layer dictates the magnetic periodicity along the *c* axis where an up-up-down-down (↑↑↓↓) spin arrangement of the in-plane zigzag AFM structure [characterized by the propagation vector ***k*** = (½ ½ ½)] is found. Our results, thus, provide a strong correlation between the magnetic moments in the transition metal layers and the Na-chiral order in the adjacent nonmagnetic Na-layers. Besides, the above described commensurate (CM) zigzag AFM order state is found to coexist with an incommensurate (ICM) AFM state below the $T_N$ ~ 27.5 K. The ICM state is found to appear at much highere temperature ~ 50 K and persists down to lowest measured temperature of 1.7 K. Our reverse Monte Carlo (RMC) analysis divulges a two dimensional (2D) magnetic correlations (within the *ab* plane) of the ICM AFM state over the entire temperature range 1.7-50 K. Further, the spin-Hamiltonian has been determined by carrying out inelastic neutron scattering experiments and subsequent linear spin-wave theory analysis which reveals the presence of competing inplane exchange interactions up to 3$^{rd}$ nearest neighbours consistent with the zigzag AFM ground state, and weak interplanar interaction as well as a weak single-ion-anisotropy. The values of the exchange constants yield that Na$_2$Ni$_2$TeO$_6$ is situated well inside the zigzag AFM phase (spans over a wide ranges of $J_2/J_1$ and $J_3/J_1$ values) in the theoretical phase diagram. The present study, thus, provides a detailed microscopic understanding of the magnetic correlations and divulges the intertwining magneto-structural correlations.

## I. INTRODUCTION:

The two-dimensional (2D) layered spin systems are focus of theoretical and experimental investigations due to their novel properties as a consequence of the enhanced role of quantum fluctuations in reduced dimensions, such as, topological phase transition [1-5], novel spin liquids ground states without long-range magnetic ordering [6], quantum Hall effect [7], and high-temperature superconductivity [8]. In particular, the 2D spin systems involving spin frustrations due to competing magnetic interactions and anisotropy play herein an important role. A striking example of such systems is layered compounds with a honeycomb lattice as the magnetic layers. The honeycomb lattice possesses the strongest quantum spin fluctuations among 2D spin systems due to the lowest coordination number. Unlike other 2D triangular or Kagome lattices, the 2D honeycomb lattice with nearest-neighbor exchange only interaction does not involve geometric frustrations. The honeycomb lattice can have geometrical frustrations in the presence of competing exchange interactions beyond the first nearest neighbors. The combined effects of the geometrical frustration and the reduced dimensionality can show various exotic ordered as well as disordered magnetic ground states as predicted theoretically for the $J_1$–$J_2$–$J_3$ model on honeycomb lattice [9-13]. Further, several exotic magnetic phenomena are reported based on the 2D honeycomb lattice antiferromagnetic (AFM) systems [14, 15]

The quasi- 2D honeycomb lattice systems are of special interest as the magnetic symmetry can be decided/tuned not only by the symmetry of the in-plane magnetic layers but also by the intermediate nonmagnetic layers which provide the 3D coupling between magnetic layers. In the context of layered honeycomb lattice, the newly discovered layered battery materials with the general formula $A^{1+}_2M^{2+}_2X^{6+}O_6$ and $A^{1+}_3M^{2+}_2X^{5+}O_6$ ($A$ = Li, Na, and K; $M$=Mn, Fe, Co, Ni, and Cu; and $X$=Te, Sb, and Bi) are of recent interest [16-28]. The compounds represent layered crystal structures formed by alternating magnetic honeycomb layers and alkali metal ionic layers. The structure of the individual honeycomb magnetic layer is formed by the mixed edge-sharing $(X/M)O_6$ octahedra in each layer and create a unique $X$-centered $MO_6$ honeycomb lattice. The alkali metal $A$-ions that are sandwiched between the honeycomb transition metal oxide layers act as a nonmagnetic separator to provide a quasi-2D magnetic structure with possible tunability of the interplanar exchange coupling. Further, the partial occupancies, disorders, and vacancies in the intermediate alkali metal layers can lead to a certain softness of the crystal structure in the perpendicular layers packing direction, therefore, provide the possibility to stack the honeycomb layers against each other in different ways. The wide variety of magnetic structures and relevant magnetic properties of these honeycomb layered oxides $A^{1+}_2M^{2+}_2X^{6+}O_6$ and $A^{1+}_3M^{2+}_2X^{5+}O_6$ is, thus, largely caused by different relative arrangements of the magnetic honeycomb layers, degree of inter layer ordering, the presence of stacking faults and their concentration, various types of alkali-metal coordination, and the distances between the layers. Evidently, spin structure types and magnetic properties of these honeycomb layered oxides are closely related to their crystal structures. Wide variations of the Néel temperature, (mostly below 40 K) depending on the $A$, $M$, and $X$ ions, was reported for these compounds [15]. Another interesting feature is the manner in which the antiparallel spins align in the honeycomb planes *i.e.*, zigzag AFM ordering and alternating stripe-like spin patterns within the honeycomb layers. Among these compounds, the P2-type compound $Na_2Ni_2TeO_6$ is unique as it reveals a coexistence of the commensurate (CM) and incommensurate (ICM) magnetic correlations [24] and is of our present interest. The compound has attracted considerable attention in recent years from both magnetism (a model quasi-2D $S =1$ honeycomb lattice system) and battery application (exhibiting high ionic conductivity at room temperature) [16, 17, 24, 25, 28-32],

Here, by comprehensive neutron diffraction and inelastic neutron scattering studies we report a novel magnetic phenomenon in the quasi-2D layered compound $Na_2Ni_2TeO_6$ where the magnetic symmetry is controlled by the crystal structural symmetry of the intermediate nonmagnetic Na-layer. Such phenomenon is a unique feature for $Na_2Ni_2TeO_6$ with respect to other Na-ion based layered magnetic compounds, especially, $A_2M_2XO_6$ or $A_3M_2XO_6$ compounds. Although there were some reports on the magnetic properties of $Na_2Ni_2TeO_6$ in literature, however, the nature of the magnetic ground state and spin correlations remain highly debatable. Karna *et al*. [24] reported the coexistence of strong ICM and weak CM AFM orderings. Whereas, Kurbakov *et al*. [17] reported a pure single-phase CM AFM ordering having completely different symmetry. Therefore, the details of the magnetic ground state and its temperature evolution remain unclear. In the present work, by a comprehensive neutron diffraction study, we have established that the magnetic ground state consists of a coexisting 3D CM zigzag AFM and a 2D ICM spin orderings below the $T_N$. The present study also provides in-depth spin-spin correlations of the two coexisting 3D CM and 2D ICM AFM phases as a function of temperatures by performing Rietveld analysis of the magnetic Bragg peaks from the 3D CM phase, and the reverse Monte Carlo (RMC) analysis of the diffuse magnetic scattering from the 2D ICM phase, respectively. Our neutron diffraction data reveal that with decreasing temperature the 2D ICM phase appears at ~ 50 K and then coexists with the 3D CM AFM phase below the $T_N$ =27.5 K. Most remarkably, we report a novel phenomenon that the up-up-down-down ($\uparrow\uparrow\downarrow\downarrow$) magnetic symmetry of the 3D CM zigzag AFM state along the $c$ axis is dictated by the intermediate non-magnetic Na-ion layers having a chiral nuclear density distribution that alternate layer to layers. Besides, we have performed an inelastic neutron scattering (INS) study to derive the spin Hamiltonian which reveals the presence of competing inplane exchange interactions up to the 3rd nearest neighbours, and a weak interplanar interaction consistent with the observed zigzag AFM ground state. The derived values of the exchange constants reveal that the compound lies well inside the zigzag AFM phase [extended over a wide ranges of $J_2/J_1$ (from +0.5 to all –ve values) and $J_3/J_1$ (for all –ve values)] in the theoretically proposed magnetic ($J_2/J_1$ - $J_3/J_1$) phase diagram. The INS results further reveal that the $Ni^{2+}$ spins, located at the trigonally distorted oxygen octahedral environment, exhibit sizeable single-ion magnetic anisotropy ($D/J_1$~ 0.15) due to the crystal field effects. The present study provides a thorough characterisation of magnetic structure and their symmetry and temperature dependent spins-spin correlations; as well as, establishes their connection to the underline crystal structure, especially the crystal structure of the intermediate non-magnetic Na-ion layers which play an important role on the mangnetism of $Na_2Ni_2TeO_6$.

.
## II. EXPERIMENTAL DETAILS:

Polycrystalline samples of $Na_2Ni_2TeO_6$ were synthesized by the solid-state reaction method [16]. The powder x-ray diffraction pattern was recorded using a Cu $K_\alpha$ radiation at room temperature. The temperature and field-dependent dc, as well ac magnetization, and heat capacity measurements were carried out using a commercial (Cryogenic Co. Ltd., UK) physical properties measurement system (PPMS). The dc-magnetization measurements were carried out over 5-300 K in the zero-field-cooled and field-cooled conditions under several magnetic fields. The ac susceptibility measurements were carried out over 5-300 K under an ac field amplitude of 5 Oe, and a frequency of 987 Hz. Temperature-dependent zero-field heat capacity was measured by an AC calorimeter.

The temperature-dependent neutron diffraction measurements were performed by using the powder diffractometers, PD-II ($\lambda$ = 1.2443 Å) at Dhruva reactor, Bhabha Atomic Research Centre, India (to derive crystal structural correlations); the DMC diffractometer ($\lambda$ = 2.4586 Å) at the Paul Scherrer Institute (PSI), Switzerland, (to determine magnetic correlations over the wide temperature range), and the E6 diffractometer ($\lambda$ = 2.40 Å), at Helmholtz Zentrum Berlin, Germany (focusing on the temperature range around the $T_N$). The measured diffraction patterns were analyzed by using the Rietveld refinement technique (by employing the FULLPROF computer program [33]). Diffuse magnetic neutron scattering spectra were analysed by Reverse Monte carlo (RMC) methos by using the *SPINVERT* computer program [34].

The inelastic neutron scattering (INS) measurements were performed on the high-flux neutron time-of-flight instrument MAPS at the ISIS facility of the Rutherford Appleton Laboratory, Didcot, United Kingdom. The INS spectra were recorded at 10, 50, and 100 K with incident neutron energy of 40 meV. Each of the INS patterns were measured for ~ 6 hours (1000 μA of incident beam). The large detector banks of MAPS spectrometer allow a simultaneous measurement over large-momentum ($Q$) regions of $S(Q,\omega)$ space. About 20-g powder sample was used for the INS measurements. The INS data were reduced using the MANTIDPLOT software package [35]. The raw data were corrected for detector efficiency and time-independent background following standard procedures. The spin-wave simulations were carried out using the SPIN-W program [36].

### III. RESULTS AND DISCUSSION:
**A. Crystal structural correlations:**

The crystal structure of $Na_2Ni_2TeO_6$ has been investigated by the combined analysis of x-ray and neutron diffraction patterns at room temperature. The Rietveld analysis of the diffraction patterns (Figs. 1 and 2) reveals that the compound crystallizes in the hexagonal symmetry with space group $P6_3/mcm$ and the crystal symmetry remains unchanged over the entire measurement temperature range 1.7-300 K. The analysis also confirms the single-phase nature of the polycrystalline sample. The lattice parameters at room temperature are found to be $a = b = 5.1990(3)$ Å and $c = 11.1297(9)$ Å. The refined atomic positions, isotropic thermal parameters, and site occupation numbers are given in Table-I. The crystal structure of $Na_2Ni_2TeO_6$ consists of the alternating layers of magnetic $NiO_6/TeO_6$ layers and nonmagnetic Na-layers [Fig. 1(c)]. The honeycomb lattices are formed by an edge shared $NiO_6$ octahedra within the *ab* planes where the $TeO_6$ octahedron occupies at the center of the honeycomb unit [Fig. 1(d)]. The crystal structure provides exchange interaction pathways up to 3$^{rd}$ nearest neighbors. Along the *c* axis, such honeycomb layers are well separated (by ~ 5.565 Å at room temperature) by an intermediate layer of Na atoms alone [Fig. 1(c)].

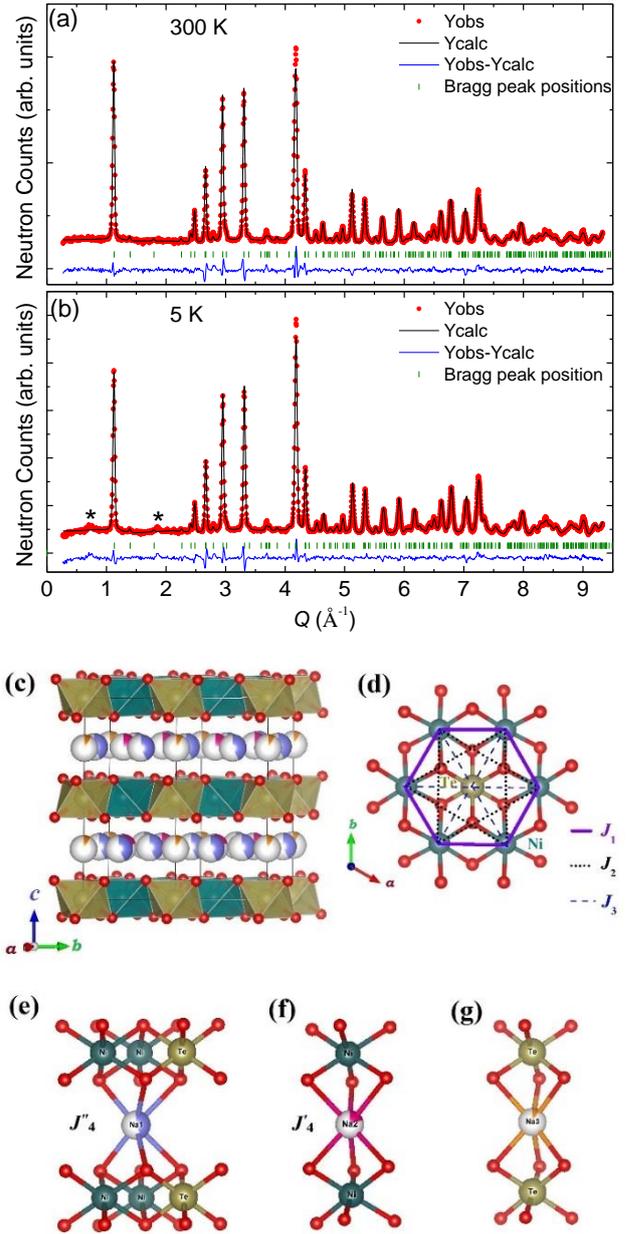

FIG. 1. (Color online) The Rietveld refined neutron diffraction patterns of $Na_2Ni_2TeO_6$ measured on PD-II, BARC, Mumbai, India, at (a) 300 K and (b) 5 K. Experimental and calculated patterns are shown by the solid circles and black lines (through the data points), respectively. The difference between observed and calculated patterns is shown by the solid (blue) lines at the bottom of each panel. The vertical bars show the allowed nuclear Bragg peaks under the hexagonal space group $P6_3/mcm$. Weak antiferromagnetic Bragg peaks ~ $Q$= 0.75 and 1.8 Å$^{-1}$ at 5 K are shown by asterisks. (c) The layered type crystal structure of $Na_2Ni_2TeO_6$. (d) A representative honeycomb unit composed of $NiO_6$ and $TeO_6$ octahedra within a given *ab* plane with possible NN, NNN, and NNNN exchange interactions $J_1$, $J_2$, and $J_3$ (for details see text). (e-g) The interlayer connections through the three Na sites. The interlayer exchange interactions $J'_1$ and $J'_2$ through the Na2 and Na1 ions are also shown.

TABLE I. The Rietveld refined atomic positions, isotropic thermal parameters, and site occupation numbers for $Na_2Ni_2TeO_6$ at room temperature.

| Atom | Site | x/a | y/b | z/c | $10^2 \times B_{iso}$ (Å$^2$) | Occ. |
|---|---|---|---|---|---|---|
| Ni | 4d | 1/3 | 2/3 | 0 | 0.34(2) | 1.0 |
| Te | 2b | 0 | 0 | 0 | 0.20(1) | 1.0 |
| O | 12k | 0.6838(5) | 0.6838(5) | 0.5930(1) | 0.65(1) | 1.0 |
| Na1 | 6g | 0.3849(1) | 0 | 1/4 | 1.25(3) | 0.42(1) |
| Na2 | 4c | 1/3 | 2/3 | 1/4 | 1.25(3) | 0.20(1) |
| Na3 | 2a | 0 | 0 | 1/4 | 1.25(3) | 0.08(1) |

In the present crystal structure with space group $P6_3/mcm$, the Na ions are distributed at three Wyckoff sites [Na1(6g), Na2(4c), and Na3(2a)], whereas, Ni (4d), Te (2b) and O (12k) ions have single Wyckoff positions [16]. The single Wyckoff positions for the Ni, Te, and O ions result in Ni and Te octahedra involving six equal bond lengths of Ni-O [=2.058(3) Å] and Te-O [=1.942(4) Å], respectively, at room temperature. However, the octahedra are found to be distorted due to the differences in the values of bond angles. The values of O-Ni-O (~ 97°) and O-Te-O (~94°) bond angles that are directed out of the honeycomb plane are found to be larger than the bond angles lying within the plane (O-Ni-O ≈ 80°, O-Te-O ≈ 86°, respectively) (Table-II). Such octahedral distortions result in slightly compressed metal oxide layers along the *c*-axis. The distortion in the $NiO_6$ octahedron further indicates the presence of a trigonal crystal field at the magnetic Ni sites. As defined by the hexagonal symmetry, the honeycomb lattice within the layers is ideal with having equal distances between all the three 1$^{st}$ nearest neighbor (NN), among all the six 2$^{nd}$ nearest neighbor (NNN), and among all the three 3$^{rd}$ nearest neighbor (NNNN) $Ni^{2+}$ magnetic ions governing the exchange interactions $J_1$, $J_2$ and $J_3$, respectively [Fig. 1(d)]. The details of the superexchange pathways are given in Table-III. Although the distance between NN $Ni^{2+}$ ions (5.565 Å) along the *c* axis between two honeycomb layers is smaller than the distance (6.088 Å) for the NNNN exchange interaction $J_3$ within the plane, the strength of the interplanar exchange interactions $J'_4$ and $J''_4$ are expected to be weaker than that of the inplane $J_3$ due to the superexchange pathways (discussed later in details).

Now we focus on the one of the most special crystal structural feature of the present compound i.e. the structure of the intermediate nonmagnetic Na-layer that uniquely dictates the magnetic symmetry as well as magnetic correlations (presented later in details). The intermediate non-magnetic layers consist of Na ions which are having a prismatic oxygen environment. The three Na triangular prismatic sites connect two adjacent honeycomb layers in different ways [Figs. 1(e)–1(g)]. The Na1 site is located between two tetrahedral voids, formed by two $NiO_6$ and one $TeO_6$ octahedron, each one from the top and the bottom layers. The Na2 and Na3 sites are situated between two $NiO_6$ and two $TeO_6$ octahedra, respectively. Therefore, there are mainly two distinct interlayer magnetic coupling pathways via the Na1 and Na2 sites (Ni-O-Na1/Na2-O-Ni). Further, the recent study by Karna et al. [24], via an inverse Fourier transform (iFT)-assisted neutron and x-ray diffraction analyses, reported the nuclear density distribution of Na ions which reveals a two-dimensional (2D) chiral pattern of well-defined handedness in the Na layers without breaking the original 3D crystal symmetry. The analyses indicated the quintuplet splitting of Na1(a-e), the triplet splitting of Na2(a-c), and the doublet splitting of Na3(a-b) sites. The nuclear density distribution of the Na1(a-e) sites reveals a circular chiral pattern surrounding the Na2a center showing alternating handedness of the two Na-layers (counterclockwise and clockwise) within the unit cell. These hidden chirality in the Na layer is indicated by the significant broadening of the Bragg peaks with indices having ($l\neq0$), in contrast to the narrow instrumental resolution limited Bragg peaks with indices having $l$=0 [24]. Consistent with the earlier report, our present x-ray diffraction study at room temperature reveals a broadening of the Bragg peaks with indices ($l\neq0$), viz., (102), (114), (116) and (304) in addition to the resolution limited sharp Bragg peaks (100) and (300) ($l$=0) [Fig. 2]. Therefore, the 2D chiral pattern of the Na nuclear density distribution is evident for the studied sample in the present work. Such chiral pattern of the Na nuclear density distribution plays a significant role in the magnetic correlations between the honeycomb layers along the $c$ axis, i.e., the nature of the magnetic ground-state, as found in our low-temperature neutron diffraction study (discussed later). Moreover, all the three Na sites are partially occupied with different percentages of Na ions [Na1 ~ 42 %; Na2 ~ 20 %; Na3 ~ 8 %]. The partial occupations of Na ions are expected to interrupt the magnetic coupling between two honeycomb planes along the $c$ axis and as a result, a coexistence of the 2D magnetic correction may expected which is evident in our low-temperature neutron diffraction study (discussed later). The effective interlayer exchange interactions are, therefore, expected to be much weaker than intralayer exchange interactions, as revealed by the reported DFT calculations [24] as well as the present INS study (discussed later). The Rietveld refinement further reveals that there is neither site mixing of Ni and Te ions, nor between Ni/Te and Na ions in the studied sample.

TABLE II. The local crystal structural parameters; bond lengths, and bond angles at room temperature.

| Site | bond length (Å) | | bond angle (°) | | |
|---|---|---|---|---|---|
| Ni | (Ni–O) | 2.058(3) | O–Ni–O | 96.92(19) | Out-of-plane |
| | | | O–Ni–O | 79.91(16) | in-plane |
| Te | (Te–O) | 1.942(4) | O-Te-O | 94.24(20) | Out-of-plane |
| | | | O-Te-O | 85.76(18) | in-plane |
| Na1 | (Na1–O) | 2.337(6) | (O–Na1–O) | 55.96(11)/ 68.07(10)/ 78.16(12) | 86.77(11)/ 96.69(15) |
| Na2 | (Na2–O) | 2.492(3) | (O–Na2–O) | 76.33(9) | 88.96(11) |
| Na3 | (Na3–O) | 2.398(3) | (O–Na3–O) | 72.81(9) | 93.48(8) |

TABLE III. Possible pathways for NN, NNN, and NNNN exchange interactions $J_1$ and $J_2$ and $J_3$, respectively. The Ni...Ni direct distances, metal oxide ($M$–O) bond lengths and bond-angles for the exchange interactions $J_1$, $J_2$ and $J_3$ in $Na_2Ni_2TeO_6$ at room temperature.

| Exchange interaction | Pathways | Ni...Ni direct distance (Å) | Bond lengths (Å) | Bond angles (deg.) |
|---|---|---|---|---|
| $J_1$ | Ni–O–Ni | Ni–Ni= 3.0001(1) | Ni–O= 2.058(3) | Ni–O–Ni= 93.62(12) |
| $J_2$ | Ni–O–Ni–O–Ni /Ni–O–Te–O–Ni /Ni–O–O–Ni | Ni–Ni= 5.1963(1) | Ni–O=2.058(3) Te–O=1.942(4) O–O=2.643(3) | Ni–O–Ni= 93.62(12) O–Ni–O=79.91(16) Ni–O–Te= 97.16(1) O–Te–O= 85.76(18) |
| $J_3$ | Ni–O–Te–O–Ni | Ni–Ni = 6.088(5) | Ni–O= 2.058(3) Te–O=1.942(4) | Ni–O–Te=97.16(1) O–Te–O = 180.0(1) |
| (interlayer) $J'_c$ | Ni–O–Na2–O–Ni | Ni–Ni = 5.5620(4) | Ni–O= 2.058(3) Na2-O=2.492(3) | Ni–O–Na2= 74.67(8) O–Na2–O=88.96(7)/138.20(7) |
| $J''_c$ | Ni–O–Na1–O–Ni | Ni–Ni = 5.5620(4) | Na1-O=2.337(3) | Ni–O–Na1= 93.86(8)/88.06(5)/ O–Na1–O= 86.77(11)/96.69(15)/145.22(9) |

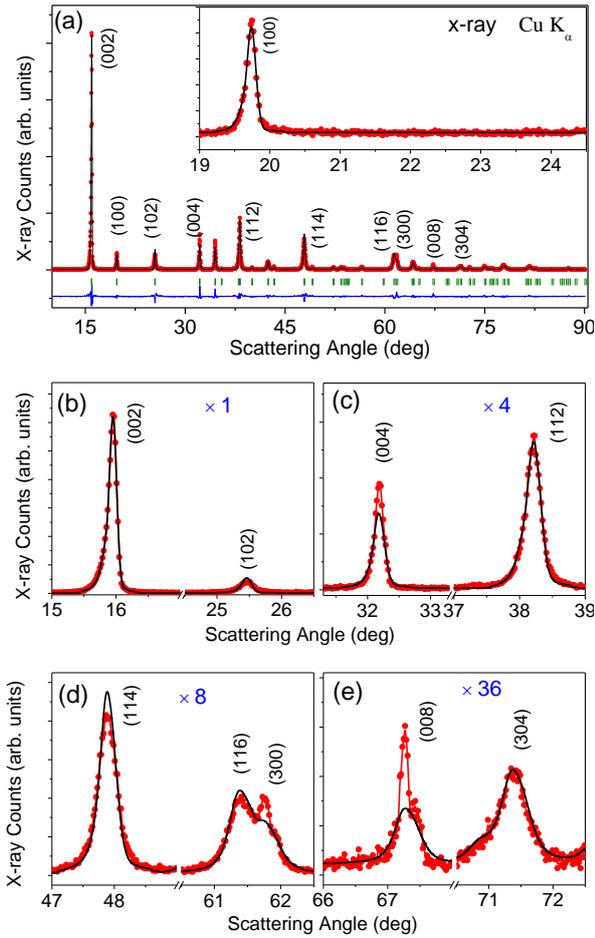

FIG. 2. (Color online) (a) The x-ray diffraction pattern of $Na_2Ni_2TeO_6$ at room temperature. Solid circles represent the experimental data points and the black lines (through the data points) are calculated pattern by Reitveld method with the average crystal structure with the space group $P6_3/mcm$. (b-e) The profile of some of the selected Bragg peaks. The Y-axis scale for (c), (d) and (e) is zoomed 4, 8, and 36 times, respectively.

Now we discuss the alternative possibility of the peak broadenings viz., due to the stacking faults in a layered crystal structure. The presence of such stacking faults is reported by Kurbakov *et al.* [17], for the studied compound $Na_2Ni_2TeO_6$, where the broadening of (002), (004), and (116) Bragg peaks with large $l$ values was reported. In contrast, our pattern reveals that the (00$l$) Bragg peaks viz., (002), (004), and (008) are sharp and resolution limited [Fig. 2]. Moreover, in contrast to the report by Kurbakov *et al.* [17], neither the characteristic "tail" near the (100) Bragg peak nor the additional Bragg peaks (101) and (103) (that are not indexed with the $P6_3/mcm$ space group), the signature of Na+/vacancy ordering, are present in our pattern. Based on the above observations it may be concluded that the sample used by us in the present study does not contain observable stacking faults, rather, contain the intrinsic chirality in the Na layers. It is also concluded that the quality of the sample reported by Kurbakov *et al.* [17] is significantly different than that used by us in the present study as well as reported by Karna *et al.*, [24]. Such difference in the sample quality leads to completely different magnetic ground states as outlined below in next sections.

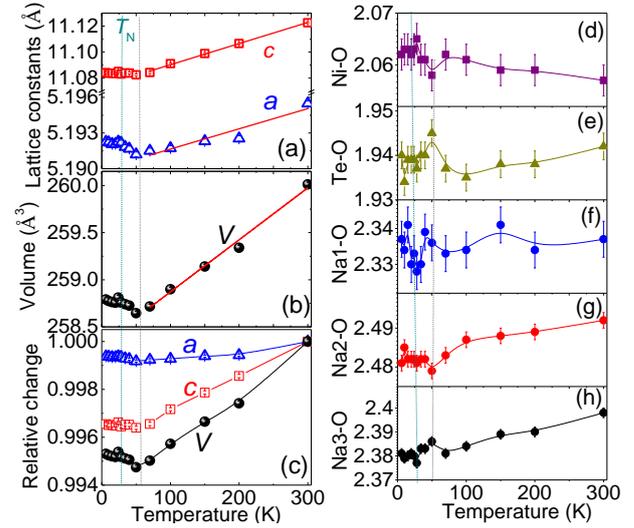

FIG. 3. (Color online) The temperature dependence of (a) lattice constants and (b) unit cell volume of $Na_2Ni_2TeO_6$. The solid lines are the linear fit to the experimental data. (c) The relative change of lattice constant and unit cell volume w. r. to the value at 300 K as a function of temperature. (d-h) The temperature-dependent metal-oxygen bond lengths.

Our temperature-dependent neutron diffraction study shows no structural phase transition or structural symmetry change down to 1.5 K. With decreasing temperature, the temperature-dependent lattice parameters $a$ and $c$ [Figs. 3(a)] show a monotonous decrease down to ~50 K. Below 50 K, the lattice parameter $c$ becomes almost constant, however, the value of the lattice parameter $a$ increases slightly with the decreasing temperature down to $T_N$~ 27.5 K and then becomes almost constant. Such anomalies below ~ 50 K are further evident in the temperature-dependent unit cell volume curve [Fig. 3(b)]. Such anomalies suggest a magneto-structural correlation. The relative changes of the lattice parameters and unit cell volume are shown in Fig. 3(c). An anisotropic thermal expansion with $\alpha_c/\alpha_a \approx 3$ is evident which is in good agreement with the earlier reports on the studied compound $Na_2Ni_2TeO_6$ [16, 24]. The anisotropic thermal expansion in $Na_2Ni_2TeO_6$ was assigned to the considerably weaker interlayer bonding than that of the intralayer bonding of a layered compound. The temperature variation of the metal-oxide bond lengths [Fig. 3(d-g)] also shows anomalies below 50 K.

**B. Bulk magnetic properties:**

The temperature-dependent dc susceptibility curve, measured under 1 Tesla of the magnetic field is shown in Fig. 4(a). The nature of the susceptibility curve is in good agreement with that reported recently by Sankar et al. [25]. The derivative curve ($d\chi_{dc}/dT$) shows a peak at ~27.5 K corresponding to the 3D long-range magnetic ordering. In order to determine the exact magnetic ordering temperature, we have also performed ac susceptibility ($\chi_{ac}$) and heat capacity measurements under zero magnetic field. The curves [Fig. 4(b)] demonstrate the magnetic long-range ordering temperature of $T_N$~27.5 K. Interestingly, the temperature-dependent susceptibility curves show a broad maximum centered around 35 K. The board peak appears due to a short-range magnetic ordering above the $T_N$~27.5 K with a possible two-dimensional magnetic correlation. Furthermore, the $\chi_{dc}T$ vs $T$ curve (upper inset of Fig. 4(a)) deviates from a constant value below ~ 200 K suggesting that the magnetic correlations start to build up below ~ 200 K.

The high-temperature $\chi_{dc}(T)$ data above 200 K are fitted with the following equation

$$\chi = \chi_0 + \frac{C}{(T - \theta_{CW})}$$

where $\chi_0$ is a temperature-independent term that accounts for the diamagnetic and Van Vleck contributions, $C$ is the Curie constant, and $\theta_{CW}$ is the Weiss temperature. The best fit gives $\chi_0$ = -30(3)×10$^{-6}$ emu/mol-Ni, $\theta_{CW}$ = -18.62(7) K, and $C$ = 2.622(1) emu-K/mol-Ni-Oe. The effective moment is estimated to be $\mu_{eff}$ = 3.24 $\mu_B$/Ni. The derived $\mu_{eff}$ value is in good agreement with the values 3.446 and 3.2 $\mu_B$/Ni reported for $Ni^{2+}$ ($S$=1) ion by Sankar et al. [25] as well as in the textbook by C. Kittle [37], respectively.

To estimate the strength of the exchange interaction we have fitted the susceptibility curve with high-temperature series expansion (HTSE) model for a 2D planar honeycomb lattice, with nearest neighbor exchange interactions only, following the approximation of Rushbrook and Wood [38] as

$$\chi = (Ng^2\mu_B^2/3kT)(S(S+1)(1 + Ax + Bx^2 + Cx^3 + Dx^4 + Ex^5 + Fx^6)^{-1}$$

where $x=J/kT$, $k = 1.3807 \times 10^{-16}$ ergK$^{-1}$, $N$ is Avogadro's number, $\mu_B =9.274 \times 10^{-21}$ erg G$^{-1}$, $g$ is the Lande-$g$ factor, $A = 4$, $B = 7.333$, $C = 7.111$, $D = -5.703$, $E = -22.281$, and $F = 51.737$ [38]. A good fit to the high temperature experimental $\chi_{dc}(T)$ data (~40–300 K) was obtained with two fitting parameters, as shown in Fig. 4(a), yielding $J/k = -8.52(6)$ K and $g = 2.05$. The value of the exchange constant is in good agreement with the value reported by Kurbakov et al.[17] as well as that determined from our inelastic neutron scattering study (discuss later in the Section E).

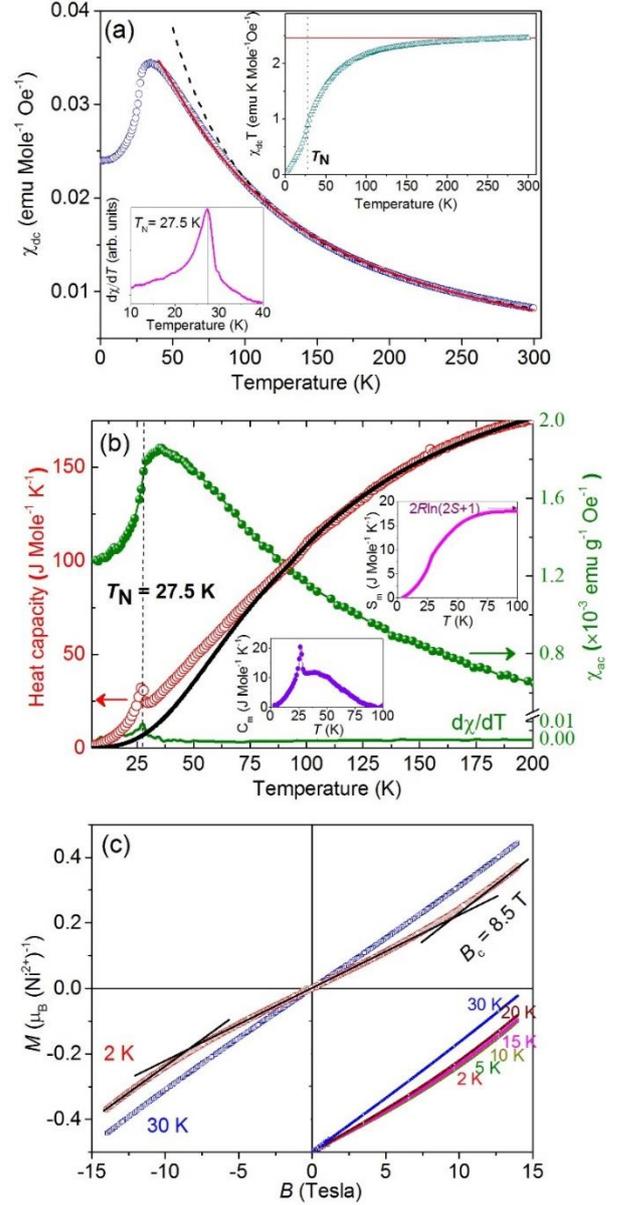

FIG. 4. (Color online) (a) The temperature dependent dc susceptibility ($\chi_{dc}$ = M/H) curves measured under an applied magnetic field of 1 T. The solid and dashed lines are the fitted curves as per the HTSE and Curie-Weiss formula (for details see the text). The top inset shows the $\chi_{dc}T$ vs $T$ curve. The red horizontal line is a guide to the eyes. The bottom inset shows the derivative curve ($d\chi_{dc}/dT$) as a function of temperature. (b) The temperature-dependent ac susceptibility ($\chi_{ac}$) and experimental heat capacity ($C_p$) curve (open symbol) for $Na_2Ni_2TeO_6$. The derivative ($d\chi_{ac}/dT$) curve is shown at the bottom. The solid black curve reveals the lattice specific heat. The insets show the temperature dependence of the magnetc heat capacity (bottom) and magnetic entropy change (top), respectively. (c) The isothermal magnetization of $Na_2Ni_2TeO_6$ as a function of the magnetic field measured at 2 and 30 K. The solid black straight lines are the guide to the eye. The inset shows the temperature variation of the magnetization curves over 2-30 K.

In order to estimate the magnetic contribution ($C_m$) to the heat capacity, we first approximate the lattice contribution ($C_{lattice}$) [shown by the solid black curve in Fig. 4(b)] by fitting the experimentally measured heat capacity curve (above 80 K) with a combination of the Debye and Einstein models of lattice heat capacity [39]. The magnetic part of the heat capacity $C_m$ is obtained by subtracting the lattice contribution from the experimentally measured data. The temperature dependent $C_m$ curve is shown in the bottom inset of Fig. 4(b). Apart from the $\lambda$-like peak due to the 3D magnetic transition at $T_N = 27.5$ K, a strong broad peak is present due to the 2D short-range magnetic ordering. The magnetic entropy, $S_m$, (deduced from the temperature integration of $C_m/T$) saturated above 90 K to a value of ~17.9 J mol$^{-1}$ K$^{-1}$ [top inset of Fig. 4(b)]. The saturation value is about 98% of the theoretical magnetic entropy of $S_m = 2R \ln(2S+1)$ of ~ 18.27 J mol$^{-1}$ K$^{-1}$. On the other hand, the entropy gain from the 3D long-range ordering below the $T_N$ is ~ 45% of the total $S_m$. The derives values of the magnetic entropy are in good agreement with that reported by Sankar et al. [25]. The significant amount of magnetic entropy gain above the $T_N$ indicates the presence of 2D short-range spin correlations.

The isothermal magnetization curves of Na$_2$Ni$_2$TeO$_6$ measured at 2 and 30 K are shown in Fig. 4(c). At 30 K (above $T_N$ ~ 27.5 K), the $M(B)$ curve shows a linear behavior, whereas, at 2 K, (in the ordered AFM state; $T<T_N$ ~ 27.5 K), the $M(B)$ curve shows a slope change (an onset of upturn) at ~ 8.5 T. With increasing temperature, the anomaly gradually becomes broad and disappears at $T > T_N$ [inset of Fig. 4(c)], confirming its relation to the ordered magnetic state. The upturn in the $M(B)$ curve suggests a field-induced spin-flop like transition and suggesting the presence of a weak anisotropy. Such a field-induced transition was reported for several other honeycomb antiferromagnets with Ni$^{2+}$ magnetic ions, viz., Na$_3$Ni$_2$SbO$_6$ [40, 41], Li$_3$Ni$_2$SbO [41], and Na$_3$Ni$_2$BiO$_6$ [42]. For all these compounds, the magnetic ground state is found to be an inplane zigzag AFM state, i.e., alternating ferromagnetic chains coupled antiferromagnetically within the honeycomb plane. For the present compound, no hysteresis is observed in the $M(B)$ curves down to the lowest measured temperature of 2 K. It is also noted that the maximum value of magnetic moment $M \approx 0.4$ $\mu_B$/Ni$^{2+}$ at the highest applied magnetic field of 14 T is only about 20 % of the theoretically expected saturation magnetic moment of 2 $\mu_B$/Ni$^{2+}$ ($M_S = gS\mu_B$/Ni$^{2+}$ = 2 $\mu_B$/Ni$^{2+}$ with $g = 2$) indicating that a much higher field is required to obtain the field polarized state.

### C. Magnetic ground state:

Now we present the central result of the present study i.e., the microscopic spin-spin correlations. To understand the microscopic nature of spin-spin correlations, we have carried out a comprehensive neutron diffraction study with fine temperature steps. Figure 5 shows the temperature-dependent neutron diffraction patterns measured over a wide temperature range both below and above the $T_N$ ~ 27.5 K. With decreasing temperature from 100 K, a broad satellite magnetic peak, centered at the scattering angle $2\theta = 16.2$ deg. ($Q=0.7$ Å$^{-1}$) starts to appear below a temperature $\approx 50$ K [Fig. 5(a)] revealing the onset of the short-range AFM correlations. With decreasing temperature, the peak becomes intense and sharper down to ~ 28 K. With further decreasing temperature, the broad peak becomes much narrower and intense below the $T_N = 27.5$ K, and an additional magnetic Bragg peak appears at the scattering angle $2\theta = 17.2$ deg ($Q=0.75$ Å$^{-1}$) [Fig. 5(b)] which becomes the most intense magnetic Bragg peak at low temperatures. The appearance of the magnetic Bragg peaks below the $T_N$ suggests the onset of the long-range AFM ordering. A detailed study with fine temperature steps around the $T_N$ ~ 27.5 K [Fig. 5(c)] reveals the temperature evaluation of the closely spacing two magnetic peaks. A small differences between the patterns in Figs 5(a-b) and Fig. 5(c) appear due to the differences in the resolution and background of the two instruments; DMC and E2 diffractometers, respectively. The positions of the magnetic peaks correspond to incommensurate (ICM) and commensurate (CM) magnetic correlations as shown by the vertical dotted and dashed lines, respectively. The magnetic signal for all the patterns is asymmetric with a long tail at the higher scattering angles. Further analyses reveal that the ICM peak at $2\theta = 16.2$ deg. is broad asymmetric with sawtooth type peak profile, whereas, the CM peak at $2\theta = 17.2$ deg. is sharp and symmetric [Fig. 6]. We discuss below first the magnetic correlations below the $T_N$ ~ 27.5 K.

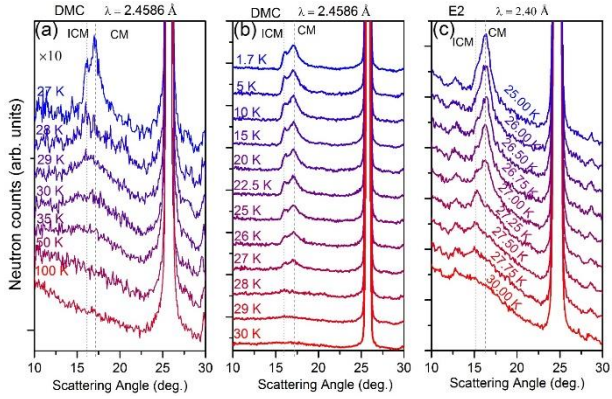

FIG. 5. (Color online) The temperature-dependent selected area of the neutron diffraction patterns of $Na_2Ni_2TeO_6$ measured over 1.7-100 K by using the neutron diffractometers (a,b) DMC, PSI, Switzerland and (c) E6, HZB, Berlin. The panel (a) highlights the diffraction patterns above the $T_N$, while, the panels (b) and (c) highlight the neutron diffraction patterns below the $T_N$. The patterns in (a) are zoomed 10 times with respect to that are shown in (b). The dashed and dotted vertical lines represent the magnetic peaks at the commensurate and incommensurate positions.

The magnetic ordering and spin structure of $Na_2Ni_2TeO_6$ below the $T_N$ were reported previously by Karna et al. [24] and Kurbakov et al. [17]. However, the results are contradictory to each other. The report by Karna et al. revealed signatures of both strong ICM [propagation vector $k$ = (0.47 0.44 0.28)] and weak CM [propagation vector $k$ = (½ 0 0)] AFM spin orderings. On the other hand, the magnetic neutron diffraction pattern reported by Kurbakov et al. [17] is completely different and shows a pure single-phase AFM ordering with the CM propagation vector $k$ = (½ 0 0). The commensurate spin structure with a single propagation vector is similar to the related $Na_2Co_2TeO_6$ compound with space group $P6_222$ [21]. The magnetic neutron diffraction patterns for our sample are close to that reported by Karna et al. [24]. Figure 6(a) compares the experimental magnetic diffraction pattern of the present sample at 1.7 K, obtained after subtraction of the nuclear background at 50 K, with the calculated magnetic patterns for the magnetic structures that are reported by Karna et al. [24] (solid line) and Kurbakov et al. [17] (dashed line), respectively. The pictorial representations of the magnetic structures are shown in [Figs. 6(e-f)] and [Figs. 6(g-h)], respectively. The calculated pattern for the magnetic structure [Figs. 6(e-f)] reported by Kurbakov et al. [17] (dashed line) completely different from our experimental pattern. On the other hand, the calculated position of the first magnetic Bragg peak indexed as (½,0,0) by Karna et al. [24] does not match with the experimental peak position (inset of Fig. 6(a)]. Such mismatch is evident for other magnetic Bragg peaks, i.e., (½ 1 0) and (½ 2 0) as well. Therefore, we rule out the possibility of the magnetic propagation vector $k$ =(½ 0 0) for the present compound. Rather, our analyses reveal that the propagation vector is $k$ = (½ ½ ½) which indexes all the magnetic peaks except the first ICM magnetic peak at $2\theta$ = 16.2 deg.

To determine the symmetry-allowed magnetic structure of $Na_2Ni_2TeO_6$, we performed a representation analysis using the program BASIREPS from the FULLPROF package [33]. The symmetry analysis for the propagation vector $k$ = (½ ½ ½) and the space group $P6_3/mcm$ gives two nonzero irreducible representations ($\Gamma_1$ and $\Gamma_2$), hence, two possible magnetic structures. Both the $\Gamma$s are two dimensional and appear three times in the magnetic representation. It results in six basis vectors for both of the representations. The basis vectors for two $\Gamma$s are given in Table –IV. Out of two $\Gamma$s, the best refinement of the magnetic diffraction pattern is obtained for the $\Gamma_1$. The Rietveld refined pattern with the magnetic propagation vector $k$ = (½ ½ ½) is shown by the solid black line [Fig. 6(b)]. The corresponding magnetic structure is shown in Figs. 6(i) and 6(j). The magnetic structure corresponds to an in-plane zigzag AFM ordering with the ferromagnetic chains running along the diagonal [110] direction. The magnetic moments are found to be pointing along the $c$ axis. Such magnetic layers are arranged in a (↑↑↓↓) UUDD fashion along the $c$ axis.

The observed UUDD (↑↑↓↓) structure is the most important result and an unique finding of the present study. We would like to point out that the found UUDD spin arrangement along the $c$ axis is unique concerning all the equivalent magnetic honeycomb layers constituted by $NiO_6$ and $TeO_6$ (Fig. 1). The coupling between the magnetic honeycomb layers along the $c$ axis occurs through two exchange interactions $J'_4$ and $J''_4$ involving the superexchange interaction pathways Ni–O–Na2–O–Ni and Ni–O–Na1–O–Ni, respectively (Fig. 7 and Table-III). For all the layers, the superexchange interaction pathways (Ni–O–Na2–O–Ni and Ni–O–Na1–O–Ni) are identical and involving the same bond lengths and bond angles (Table-III). Therefore, the spin arrangements between nearest layers are expected to be uniform, i.e., either UDUD ((↑↓↑↓)) or UUUU (↑↑↑↑) or DDDD (↓↓↓↓) . In contrast, our results reveal a UUDD (↑↑↓↓) spin arrangement along the $c$ axis (Fig. 6). This is to be noted that along the $c$ axis the two neighboring Na-layers have opposite chirality in the nuclear density distributions (Fig. 7) [24]. Therefore, the observed double periodicity of the magnetic spin arrangement along the $c$ axis (UUDD spin arrangement) reveals that the change of the sign of the magnetic moment occurs for a particular type of chirality of the Na-ion layer (Fig. 7). It is found that the change of the sign of the magnetic moment occurs only when the chirality is 'left (L)', whereas, no change of sign of magnetic moments when the chirality is 'right (R)'. This is a rare phenomenon where the magnetic symmetry is dictated by the crystal structure of the intermediate nonmagnetic layer.

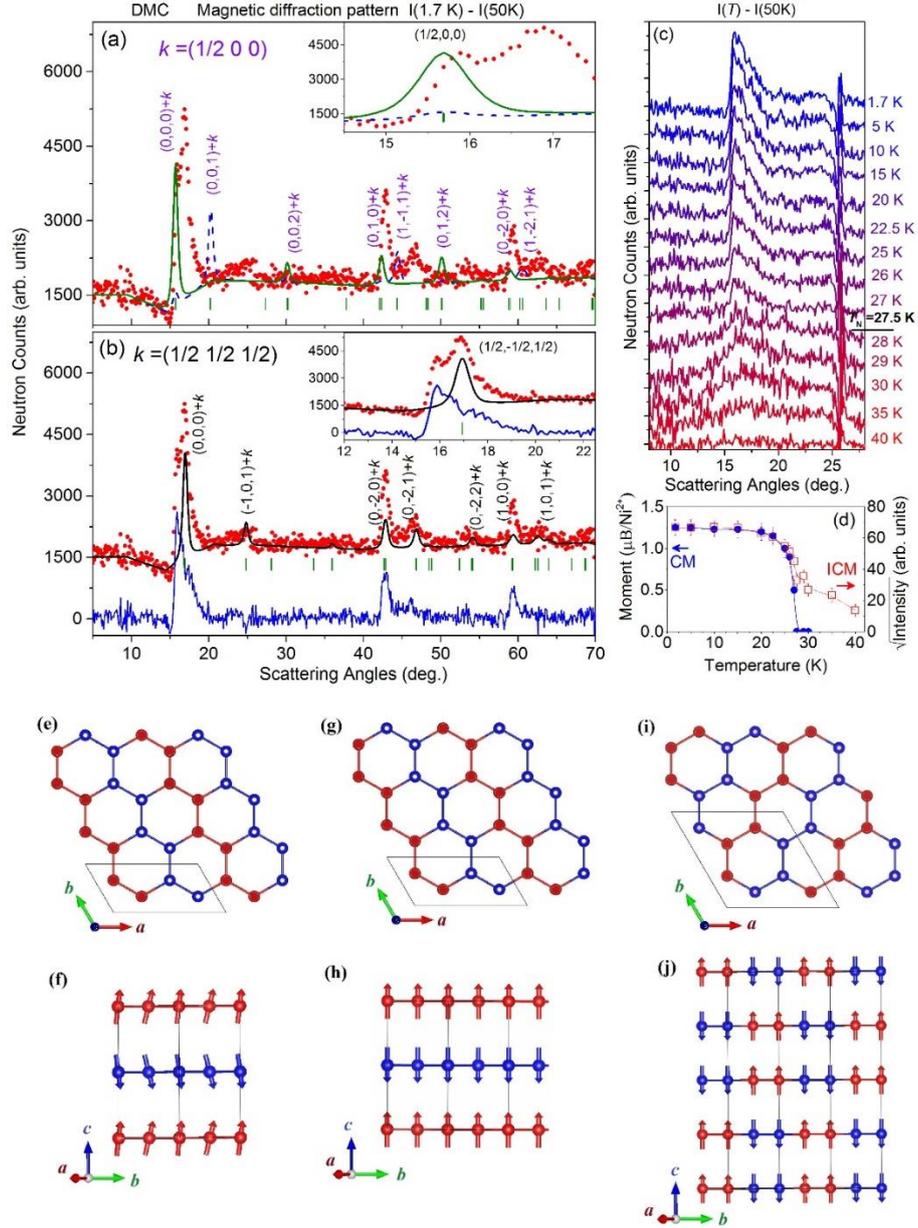

FIG. 6. (Color online) (a) The magnetic diffraction pattern (data points) $Na_2Ni_2TeO_6$ at 1.7 K, measured on DMC diffractometer ($\lambda$ = 2.5486 Å), after subtraction of nuclear background at 50 K. The calculated magnetic diffraction patterns as per the ref. [24] (solid line) (the magnetic structure shown in (g) and (h)], and ref. [17] (dashed line) (the magnetic structure shown in (e) and (f)], respectively. (b) The calculated magnetic diffraction pattern as per the magnetic structure, as shown in (i) and (j), was determined in the present study along with the experimental magnetic pattern. The insets in (a) and (b) show the enlarged views of the diffraction patterns. (c) The temperature evolution of the asymmetric incommensurate magnetic peak. The patterns above the $T_N$ =27.5 K are obtained by subtraction of nuclear background at 50 K. The patterns bellow the $T_N$ are considered as the difference patterns of the refinements of the magnetic patterns by the commensurate magnetic structure with the propagation vector $k$ = (½ ½ ½). Such a difference pattern for 1.7 K is shown by the blue line at the bottom of the panel (b). The magnetic structures (e-f) reported by Kurbakov *et al.* [17], (g-h) reported by Karna *et al.* [24], and (i-j) determined in the present study.

The observed inplane zigzag AFM structure of honeycomb lattices (within the *ab* planes) cannot be explained by the NN exchange interaction $J_1$ alone in which case the ground state is a non-frustrated Néel type antiferromagnet. The collinear zigzag AFM state in a honeycomb lattice is a result of the "order-by-disorder" phenomenon. As reported by several theoretical studies [9, 13], the zigzag AFM ground state in a honeycomb lattice, however, is possible in the presence of competing NN, NNN, and NNNN interactions $J_1$, $J_2$, and $J_3$. Our inelastic neutron scattering study yields the presence of NN, NNN, and NNNN interactions in the studied compound $Na_2Ni_2TeO_6$ (discussed later).

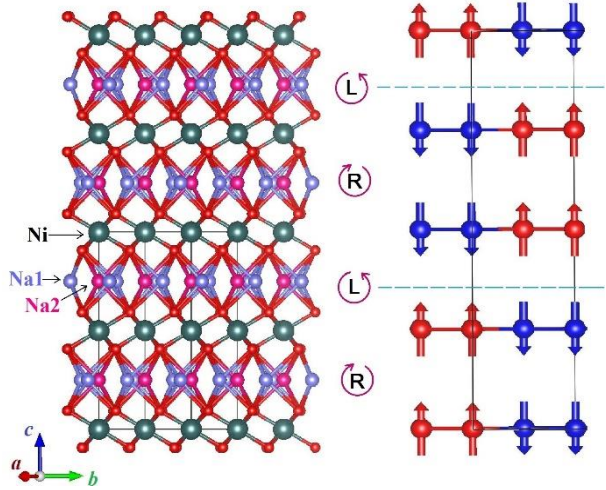

FIG. 7. (Color online) The correlation between chirality in the Na layers and the magnetic symmetry of $Na_2Ni_2TeO_6$ along the $c$ axis. The chirality of the Na-ions in the intermediate layers is shown by the right 'R' and left 'L' symbols. The change of sign of the magnetic moments (up or down) occurs only for a particular chirality, viz., for the "left (L)" chirality as shown in the figure by horizontal dashed lines, which leads to a doubling of the magnetic unit cell along the $c$ axis. The dimensions of the nuclear and magnetic cells are shown by the rectangles for each of the structures, respectively.

TABLE IV. Basis vectors of the magnetic Ni sites with $k = $ (½ ½ ½) for $Na_2Ni_2TeO_6$. Only the real component of the basis vectors are presented. The atoms within a primitive unit cell are defined according Ni-1 (0.3333 0.6667 0.0000); Ni-2 (0.6667 0.3333 0.5000); Ni-3 (0.6667 0.3333 0.0000); and Ni-4 (0.3333 0.6667 0.5000).

| IRs | | Basis Vectors | | | |
|---|---|---|---|---|---|
| | | Ni-1 | Ni-2 | Ni-3 | Ni-4 |
| $\Gamma_1$ | $\Psi_1$ | (100) | (000) | (0-10) | (000) |
| | $\Psi_2$ | (010) | (000) | (-100) | (000) |
| | $\Psi_3$ | (001) | (000) | (00-1) | (000) |
| | $\Psi_4$ | (000) | (010) | (000) | (-100) |
| | $\Psi_5$ | (000) | (100) | (000) | (0-10) |
| | $\Psi_6$ | (000) | (00-1) | (000) | (001) |
| $\Gamma_1$ | $\Psi_1$ | (100) | (000) | (010) | (000) |
| | $\Psi_2$ | (010) | (000) | (100) | (000) |
| | $\Psi_3$ | (001) | (000) | (001) | (000) |
| | $\Psi_4$ | (000) | (010) | (000) | (100) |
| | $\Psi_5$ | (000) | (100) | (000) | (010) |
| | $\Psi_6$ | (000) | (00-1) | (000) | (00-1) |

The difference pattern in Fig. 6(b) shows the magnetic contribution of the ICM phase, interestingly, consisting of three asymmetric sawtooth like peaks at $2\theta$ = 16.2, 42.7, and 59.3 degs. The peak profile of magnetic diffraction patterns depends on the dimensionality of the magnetic ordering [either 2D or 3D]. For a 2D magnetic ordering, rod-like scatterings appear in the reciprocal space as there is no restriction on the third direction. The powder averaging of such rod-like scatterings results in asymmetric sawtooth type peaks, defined by the Warren function, in the powder diffraction pattern [43-46]. On the other hand for a 3D magnetic ordering, symmetric Bragg peaks, defined by a Lorentzian function, are obtained in the powder diffraction patterns [43, 44]. The temperature dependence of the strongest asymmetric sawtooth-like peak at $2\theta$ = 16.2 deg. [Fig. 6(c)] reveals that the peak intensity is present even above the $T_N$ = 27.5 K, and persists up to ~ 50 K. Therefore, in summary, it may be concluded that with decreasing temperature an inplane 2D incommensurate magnetic correlation appears below ~ 50 K and remains 2D down to lowest measured temperature of 1.7 K. Besides, a commensurate 3D AFM ordering with propagation vector $k$ = (½ ½ ½) occurs below the $T_N$ = 27.5 K and coexists with the 2D incommensurate magnetic correlation down to lowest measured temperature 1.7 K. The observed coexistence of the CM and ICM orderings in a honeycomb lattice AFM, like the studied compound $Na_2Ni_2TeO_6$, is novel and has a unique origin (discuss later).

We would like to further comment that the microscopic magnetic properties and the magnetic ordering temperature of $Na_2Ni_2TeO_6$ are strongly dependent on both the internal crystal symmetry and the Na content. In this regard, Karna et al. [24], reported that the $T_N$ is extremely sensitive to the excess Na content, where the $T_N$ value decreases from 27.5 K to ~ 22 K when the Na content was increased from 2 to ~2.16. It is important to mention here that there are several differences in the crystal structure of the sample used in the present study and that was used by Kurbakov et al.[17] which were discussed in the previous crystal structural section. Besides, the magnetic ordering temperature of 25 K as reported by Kurbakov et al.[17] is lower than that found in the present study as well as that reported by Karna et al. [24]. Furthermore, the $dM/dT$ curve reported by Kurbakov et al.[17] contains two peaks at 25 and 27 K suggesting two magnetic transitions which are in clear contrast to the single peak at 27.5 K in the present study, as found from all dc-, ac-susceptibility, as well as specific heat curves [Fig. 4].

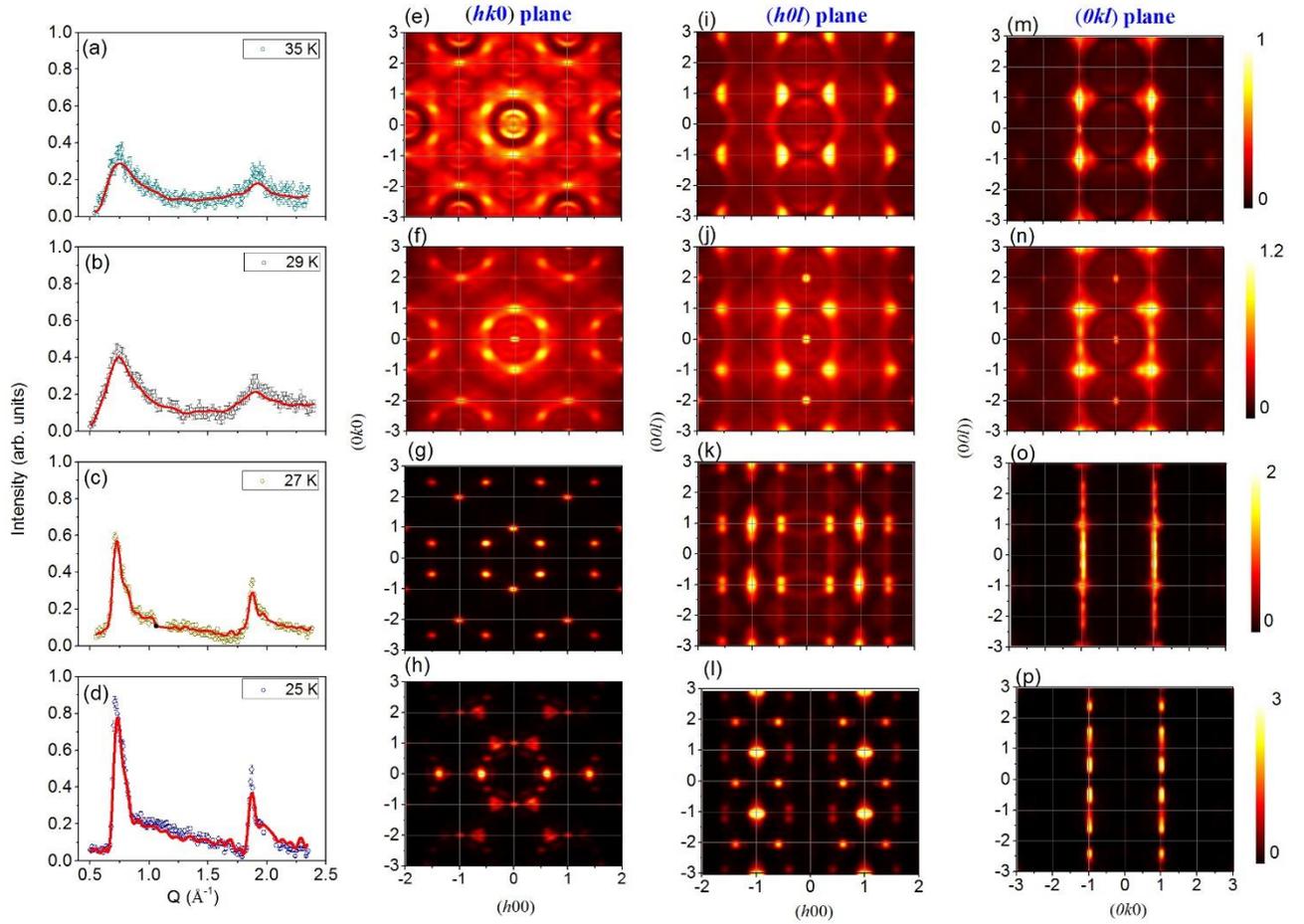

FIG 8: The experimentally measured diffuse magnetic scattering at (a) 35, (b) 29, (c) 27, and (d) 25 K The patterns are same as that shown earlier in Fig. 6(c). The diffuse magnetic scattering patterns for the temperatures above the $T_N$ =27.5 K are obtained by subtraction of nuclear background at 50 K. Whereas, the diffuse scattering patterns bellow the $T_N$ are considered as the difference patterns of the refinements of the magnetic patterns by the commensurate magnetic structure with the propagation vector $k$ = (½ ½ ½). The solid lines in each panel are the calculated scattering intensities by the RMC method. (e–p) The reconstructed diffraction patterns in the ($hk$0), ($h$0$l$), and (0$kl$) scattering planes.

## D. 2D magnetic correlations:

We now discuss the nature of spin correlation for the ICM phase that is found to be present below ~ 50 K and coexists with the 3D zigzag AFM phase below the $T_N$ ~ 27.5 K [Fig. 8]. The onset of the ICM phase at ~50 K, at a temperature almost twice the $T_N$, is consistent with the sharp decrease of the $\chi T$ values ~ 50 K [inset of Fig. 4(a)]. With decreasing temperature, the intensities of the broad peaks with the maximum at $Q$ ~ 0.7 (2$\theta$ = 16.2 deg.) and 1.9 Å$^{-1}$ (2$\theta$ = 43 deg.) increase slowly down to the $T_N$ = 27.5 K and then enhance strongly on further lowering of the temperature without any change in the peak position. This indicates that the magnetic periodicity of the ICM phase remains unchanged at temperatures above and below the $T_N$. Similar broad diffuse magnetic peaks in neutron diffraction patterns were reported for several quasi-2D layered spin systems including $Na_2Ni_2TeO_6$ [32] and the related compound $Na_2Co_2TeO_6$ [43-48]. As discussed earlier that asymmetric sawtooth type peaks, defined by the Warren function, are expected for the 2D magnetic orderings where the peak width is inversely proportional to the planar correlation length. In the present case, although the peaks are asymmetric, however, the peak shape is more complex than the simple Warren function. For a quantitative analysis of the diffuse scattering data from the ICM phase, we have employed the RMC algorithm-based SPINVERT program [34] which was successfully applied recently to several frustrated magnetic systems showing diffuse magnetic scatterings [47, 49-51]. In this program, an RMC algorithm is used to fit the experimental powder data (pure magnetic pattern) by considering a large configuration of spin vectors. There are several advantages of such an RMC method over the other model-dependent techniques (such as simple curve fitting) for the analysis of diffuse neutron scattering. The RMC method is entirely independent of a spin Hamiltonian. Therefore, it is not necessary to assume a form of the Hamiltonian to model the spin correlations. The RMC approach is superior in both quantity and accuracy of the information it provides. The only limitation of this method is that it provides a probable spin configuration out of several possibilities. This limitation can be overcome by taking an average of a large number of simulation runs. Furthermore, the SPINVERT program also calculates scattering profiles in the selected reciprocal planes by using the fitted spin configuration and the crystal structural information.

As the program SPINVERT works with orthogonal axes, we have converted the hexagonal unit cell to an equivalent orthorhombic cell having twice the number of magnetic atoms. The transformation matrix for this case is given by

$$\begin{bmatrix} a' \\ b' \\ c' \end{bmatrix} = \begin{bmatrix} 1 & 0 & 0 \\ 1 & 2 & 0 \\ 0 & 0 & 1 \end{bmatrix} \begin{bmatrix} a \\ b \\ c \end{bmatrix}$$

In the present calculations, a supercell of 30×30×20 (144000 spins) of the orthorhombic crystal structure is generated, and a randomly oriented magnetic moment is assigned to each magnetic Ni site. The positions of spins are fixed at their crystallographic sites throughout the refinement, while their orientations are refined to fit the experimental data. A total of 600 moves per spin is considered for each of the calculations.

The calculated diffuse magnetic scattering intensities are shown in Figs. 8(a-d) by the solid lines along with the experimental data (filled circles) at 35, 29, 27, and 25 K. The resulting spin configurations were used to reconstruct the $Q$ dependence of the diffuse scattering in the ($hk$0), ($h$0$l$), and (0$kl$) scattering planes [Figs. 8(e-p)] by using the SPINDIFF program extension to the SPINVERT program [34]. Above the $T_N$ = 27.5 K, rodlike diffuse scatterings along the (00$l$) direction are evident for both the ($h$0$l$) and (0$kl$) scattering planes. On the other hand, symmetric type scatterings are found within the ($hk$0) plane. The rodlike scatterings along the (00$l$) direction reveal that the magnetic correlations are confined within the 2D honeycomb planes ($ab$ plane). For a 2D magnetic ordering, as there is no restriction imposed on the $l$ value, a rodlike scattering occurs along the (00$l$) direction. Moreover, the symmetric type of scattering within the ($hk$0) plane suggests an isotropic spin-spin correlation within the honeycomb planes. With lowering of the temperature below the $T_N$= 27.5 K, the diffuse scatterings become sharp, however, retain their basic characteristics, viz., the rodlike scatterings along the (00$l$) direction and the symmetric nature in the ($hk$0) plane. This implies that the basic symmetry of the incommensurate phase remains unchanged with temperature, however, a sharp increase in the correlation lengths, especially in the $ab$ plane, occurs below the $T_N$. It is interesting to note that the 2D nature of magnetic correlations of the ICM phase remains even below the $T_N$. Similar magnetic diffuse scatterings for Na$_2$Ni$_2$TeO$_6$ were reported by Korshunov $et$ $al$., for the temperatures above the $T_N$, and 2D magnetic correlations were confirmed from an RMC analysis. The patterns [Figs. 8(k and l)] further reveal incommensurate magnetic peaks in the ($h$0$l$) scattering plane for the magnetic peaks with an index of [(2$h$+1)/2, 0, (2$l$+1)±δ]. Therefore, it is evident that the incommensurability is along the [00$l$] direction.

**E. Magnetic excitations and spin-Hamiltonian:**
The color-coded inelastic neutron scattering intensity maps of Na$_2$Ni$_2$TeO$_6$, measured on the MAPS spectrometer, ISIS facility, UK at $T$ = 10, 50, and 100 K with incident neutron energies $E_i$ = 40 meV are shown in Fig. 9. For the 10 K patterned measured within the ordered magnetic state below the $T_N$ =27.5 K [Fig. 9(a)], all the observable magnetic scatterings are situated below ~13 meV with a gap of ~ 2 meV at the AFM zone center at ~ |$Q$| = 0.7 Å$^{-1}$. The excitation intensities are mainly concentrated with two energy bands over 2-7 meV and 10-13 meV. The magnetic character of the scattering is evident from the decreasing intensity with increasing |$Q$|. The magnetic scatterings are found to be extended up to |$Q$| ~ 4.5 Å$^{-1}$. On the other hand, the patterns at 50 and 100 K, measured above the $T_N$, show gapless broad magnetic excitations which indicates the presence of short-range spin-spin correlations within the two-dimensional planes consistent with the bulk magnetization and neutron diffraction results. It is important to note that a significant amount of intensity of the magnetic excitations that has a structure in $Q$ is still present at 100 K, a temperature around four times higher than that of the $T_N$ =27.5 K. The energy dependence of the magnetic intensities, integrated over the momentum range of |$Q$| = 0- 4.5 Å$^{-1}$, is shown in Fig. 9(d). Two distinct peaks are evident at 10 K, whereas, quasi-elastic continuum scatterings are evident for 50 and 100 K. Besides, no observable phonon modes, whose intensity increases with |$Q$| as well as temperature, are evident around the spin-wave spectra over the studied momentum and energy range. This makes our data clean and easy to analyze/explain without subtracting the phonon background. The scattering cross-section $S$(|$Q$|,$\omega$) of the present polycrystalline samples, the powder average of the spin-spin correlation function $S$($Q$, $\omega$), does not carry the information regarding the direction of $Q$; however, it preserves singularities arising in the density of states as a function of $E$=$\hbar\omega$ and contains distinctive fingerprints of the spin Hamiltonian which can be readily compared with theoretical calculations to obtain approximate parameters. To model the experimentally observed magnetic spectrum, we have calculated the spin-wave dispersions, the spin-spin correlation function, and the neutron scattering cross-section using the SpinW program [36]. The studied compound $Na_2Ni_2TeO_6$ contains only the magnetic ions $Ni^{2+}$ ($3d^8$, $S = 1$), and therefore, only interactions between the $Ni^{2+}$ ions need to be considered. Considering the layered crystal structure of $Na_2Ni_2TeO_6$ with in-plane honeycomb lattice, we have constructed the magnetic Hamiltonian with exchange couplings up to third nearest neighbors [shown in Fig. 1(d)] as

$$H = \sum_i J_1(\vec{S}_i.\vec{S}_{i+1}) + J_2(\vec{S}_i.\vec{S}_{i+2}) + J_3(\vec{S}_i.\vec{S}_{i+3}) + J_4\sum_{ij}(\vec{S}_i.\vec{S}_j) + \sum_i D(S_i^z)^2$$

where $J_1$, $J_2$, and $J_3$ are the nearest neighbor, next-nearest neighbor, and next-to-next-nearest neighbor inplane exchange interactions, $J_4$ is the interplanar exchange interactions, and $D$ is the single-ion-anisotropy which originates from the crystal field of the surrounding oxygen ions in a $NiO_6$ octahedral environment. The anisotropy parameter $D$ induces an energy gap of ~ 2 meV between the ground state and excited states, as found in the experimentally measured spectrum [Fig. 9(a) and (d)].

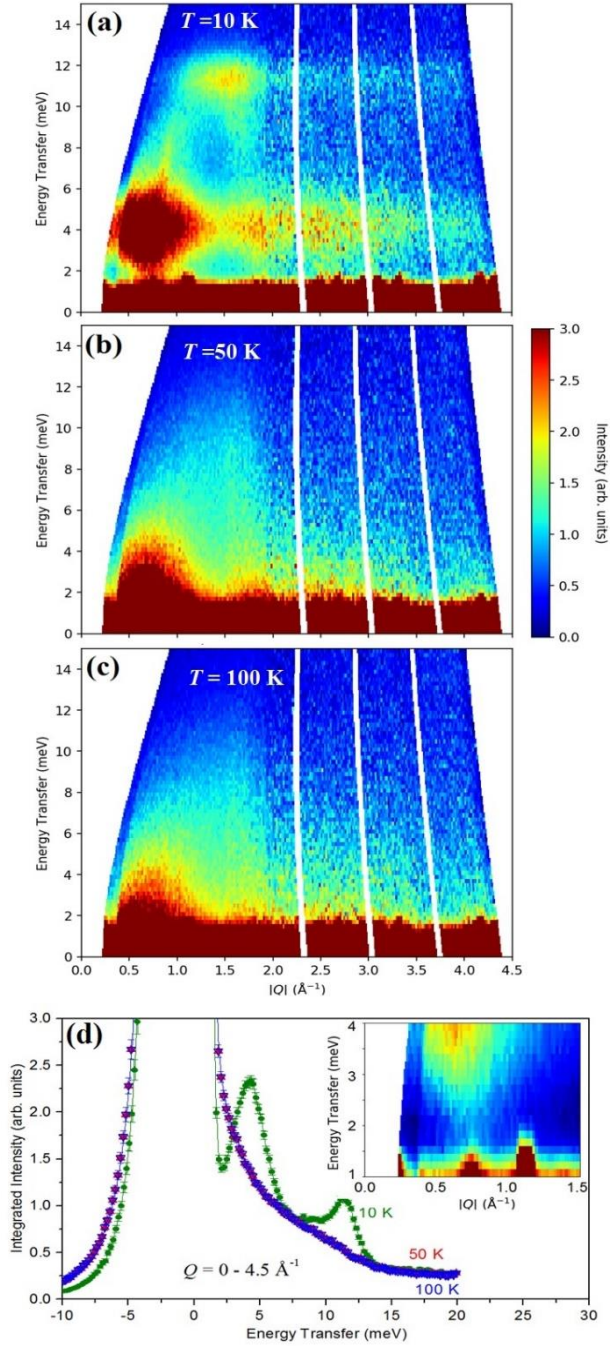

FIG. 9. (Color online) The 2D color map of the INS intensity of $Na_2Ni_2TeO_6$ as a function of energy transfer ($\hbar\omega$) and momentum transfer ($|Q|$) at (a) 10, (b) 50, and (c) 100 K, measured on the MAPS spectrometer, with incident neutron energy of $E_i = 40$ meV. The color scales show the scattering intensity $S(|Q|,\omega)$ in an arbitrary unit. (d) The intensity vs energy transfer curves at 10, 50, and 100 K. The intensities were obtained by integrations over $|Q| = 0$–4.5 Å$^{-1}$. Inset shows the selected area excitation spectrum over the lower edge revealing the energy gap and the momentum dependence of the lower edge of the lowest-energy band at 10 K.

The simulation of the spin-wave spectra is based on the CM zigzag AFM spin structure as determined in the present study [Fig. 6(h-i)] having a spin component along the $c$ axis. The spin-wave calculations assumed a magnetic form factor corresponding to $Ni^{2+}$ and a spin value $S = 1$. Solution of the Hamiltonian was tested over the wide range of parameters ($J_1$, $J_2$, $J_3$, $J_4$, and $D$) spaces. Additional details on the fitting procedure, extraction of the model solution and estimation of uncertainty are given in the Appendix-A. The tested sets of parameters are all compatible with a zigzag magnetic order as per the reported theoretical phase diagram for the $J_1$-$J_2$-$J_3$ honeycomb lattice spin system [13]. As inferred from the phase diagram, the zigzag AFM order occurs for the ferromagnetic $J_1$ and antiferromagnetic $J_2$ and $J_3$. The calculated spectra for few sets of values of the parameters are shown in Fig. 10. The possible solutions over all the parameter spaces are shown in Appendix-A (Fig. 14). A possible solution is represented by the following parameters: $J_1 = -1.40$ meV, $J_2 = 1.10$ meV, $J_3 = 1.00$ meV, $J_4 = 0.08$ meV and $D = -0.20$ meV (Table-V), and the corresponding simulated powder averaged excitation pattern is depicted in Fig. 11(b). For this solution, the $\tilde{\chi}^2_{INS}$ value is found to be 0.91. The simulated dispersion curves along the principle axes [Fig. 11(c-d)] reveal four non-degenerate dispersion modes having bandwidth between 2-12 meV within the magnetic $ab$ plane. Whereas, a weak dispersion in the simulated curves is evident along the $c$ axis indicating the weak (~1/10 times) interlayer exchange interactions $J_4$. It may be noted that the two interlayer exchange interactions $J'_c$ and $J''_c$ are indistinguishable in the present study based on the power sample. The simulated energy and momentum cuts are shown in Figs. 11(e-g) along with that obtained from the experimentally measured pattern. As seen in the figure, the model gives a satisfactory description of the main features of the magnetic excitations, except some intensities around the $\Delta E$~ 12 meV and $Q$ ~ 1.5 Å$^{-1}$. The small discrepancies in the intensity of the excitation spectra may arise from the ICM phase which is not considered in the spin-wave simulations as the nature of it's ground state is yet to be determined. Further, it may be mentioned here that such an additional intensity in the present case is unlikely due to a Kitaev interaction that was proposed recently for spin-1 honeycomb lattice system [52] (for details see Appendix-B). It may also be noted that the value of the interplanar coupling $J_4$ is relatively stronger than that reported for the related compound $Na_2Co_2TeO_6$. Such differences may arise due to the difference in the crystal structures (different space groups) of $Na_2Co_2TeO_6$ and $Na_2Ni_2TeO_6$ that leads to different stacking arrangements of the magnetic honeycomb layers along the $c$ axis. For the Ni-compound, the honeycomb lattices are stacked exactly top of each other. Whereas, for the Co compound, the neighbouring honeycomb layers have an inplane (in the $ab$ plane) shift with respect to each other. The shift is such that the centre of a hexagon match to a corner of the hexagons in the next layer. Such a shift between the magnetic layers may lead to a relatively weaker interplanar coupling in $Na_2Co_2TeO_6$ as reported recently [56]. The dominant interactions are found to be FM and operate between the NN $Ni^{2+}$ ions within the honeycomb lattice through the superexchange pathway Ni-O-Ni with a bond angle (∠ Ni-O-Ni ~94° (Table-III). As per the Goodenough-Kanamori rule [57, 58], a ferromagnetic interaction is favorable for such a superexchange interaction pathway involving an angle close to 90°. In summary, the fitting of the coupled honeycomb lattice model parameters to the experimental data reveals an essential information of strongly 2D magnetic lattice in $Na_2Ni_2TeO_6$. The derived values of the exchange constants show the presence of a strong competition between in-plane NN, NNN, and NNNN exchange interactions. The presence of the energy gap ~ 2 meV in the magnon excitation spectrum below $T_N$ = 27.5 K is consistent with the reported NMR data that showed a rapid drop of $1/^{23}T_1$ resulting from the suppression of low-energy excitations by the energy gap [30]. Our spin-wave calculations reveal a uniaxial single-ion-anisotropy with the anisotropy parameter $D = -0.20$ meV. The anisotropy axis is found to be along the $c$ axis which is the magnetic easy axis. The GGA based DFT calculations [24] also reveal a single-ion magnetocrystalline anisotropy along the $c$ axis, consistent with our experimental data. The INS data further reveal that the energy gap disappears above the $T_N$ [Fig. 9)] which is consistent with the reported NMR results [30]. A significant broadening of the excitation bands is also evident above the $T_N$ [Fig. 9].

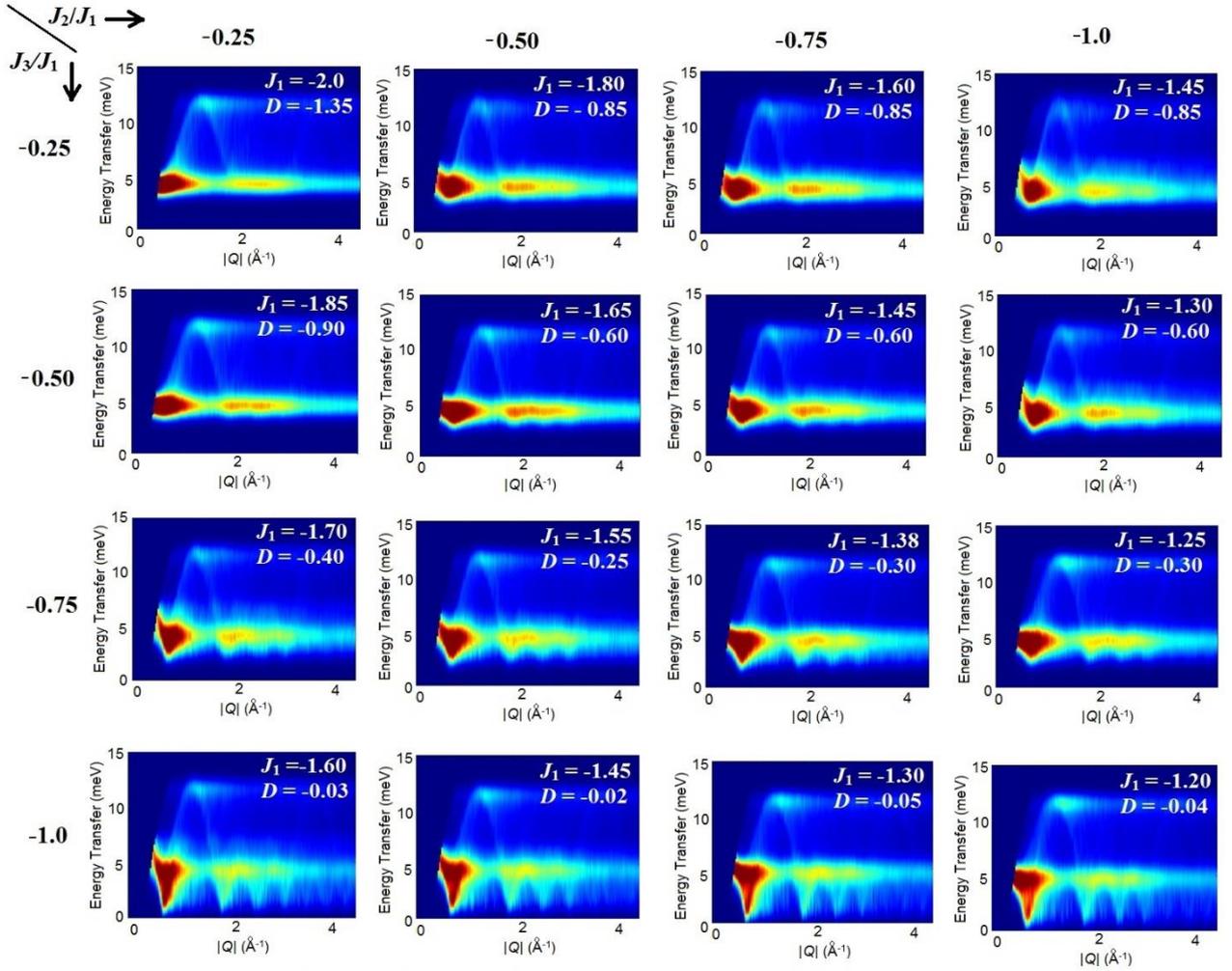

Fig. 10: Simulated (by the SPINW program) spin-wave excitation spectra as per the $J_1$-$J_2$-$J_3$ honeycomb lattice model (Hamiltonian given in text) for a series of values of $J_2/J_1$ and $J_3/J_1$ (without the interplanar coupling, i.e., $J_4 = 0$). The calculated spin-wave pattern is powder averaged, convoluted with the instrumental resolution, and corrected for the $Ni^{2+}$ magnetic form factor. For each of the cases, the values of $J_1$ and $D$ are refined to match the energy dependent two main experimental excitation peaks at ~ 4.0 and 12 meV [Figs. 9(d) and 11(e)]

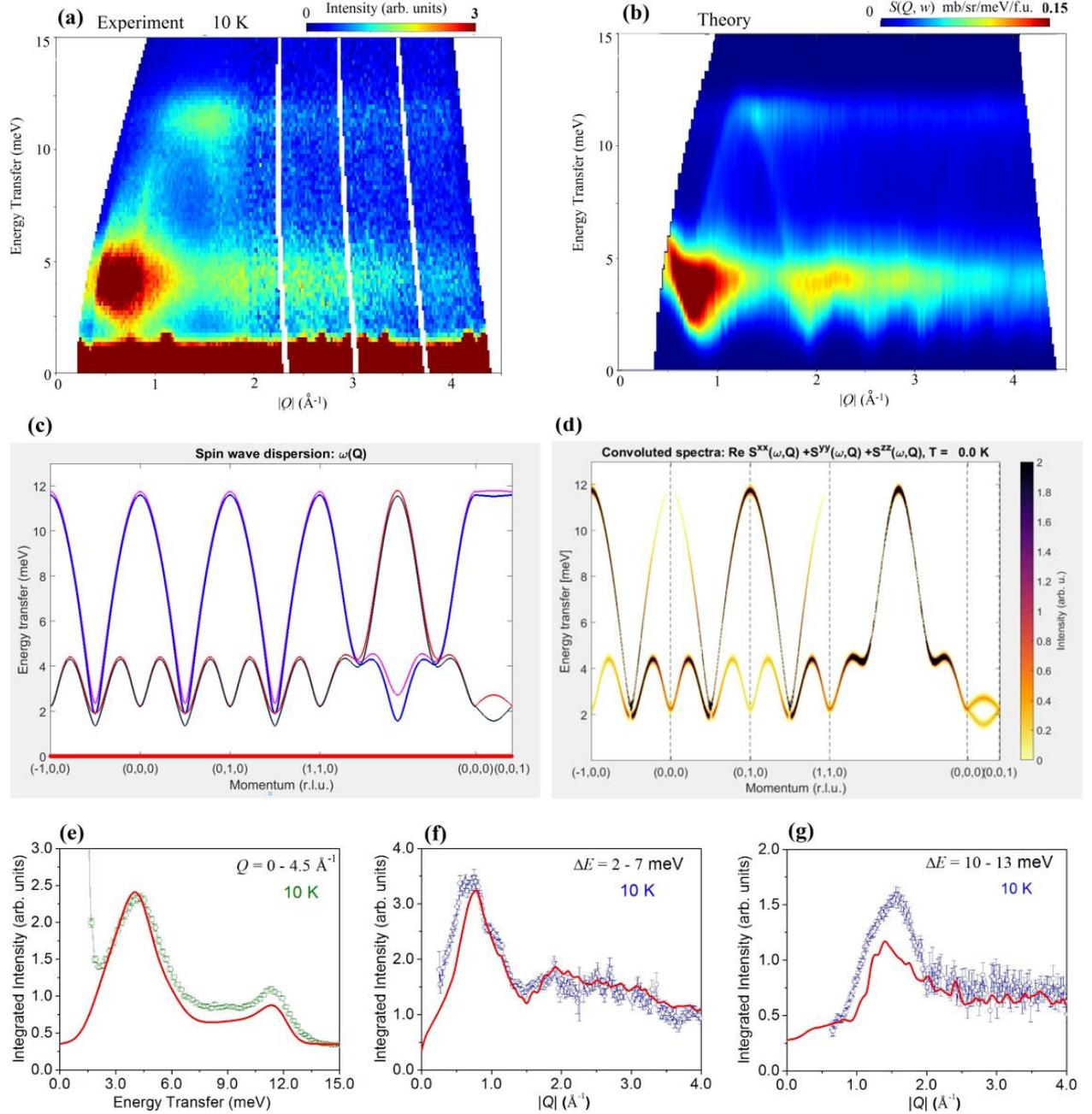

FIG. 11. (Color online) (a) Experimentally measured (at 10 K) and (b) simulated (by the SPINW program) spin-wave excitation spectra. The calculated spin-wave pattern is powder averaged, convoluted with the energy-transfer-dependent instrumental resolution, and corrected for the $Ni^{2+}$ magnetic form factor. (c) The simulated dispersion curves along the different crystallographic directions with the derived parameters $J_1 = -1.4$, $J_2 = 1.1$, $J_3 = 1.0$ and $D = -0.2$ meV. (d) The intensity variation of the dispersion patterns is shown by the color map. (e) The experimental scattering intensity as a function of energy transfer (integrated over $|Q|$ range 0–4.5 Å$^{-1}$. (f) and (g) The experimental scattering intensity as a function of momentum transfer, integrated over $\Delta E = 2$–7 meV and $\Delta E = 10$–13 meV, respectively. The spin-wave calculated intensities (red solid lines) are also plotted for comparison. To match the experimental intensity, a constant scale factor of 20 to the calculated intensity has been applied in addition to a constant background of 0.05.

TABLE V. The fitted values of the exchange interactions $J$'s and anisotropy parameter ($D$) from inelastic neutron scattering spectra at 10 K. For this solution, the $\tilde{\chi}^2_{INS}$ value is found to be 0.91. All the values of exchange interactions are in meV.

| Exchange interaction | Values (INS) (meV) | DFT-GGA (meV) [24] |
|---|---|---|
| $J_1$ | -1.40 [FM] | -0.2 |
| $J_2$ | 1.10 [AFM] | 1.2 |
| $J_3$ | 1.00 [AFM] | 0.1 |
| $J_4$ | 0.08 [AFM] | |
| $D$ | -0.20 | |

A comparison between the derived values of exchange coupling parameters from our INS data with the values reported from the DFT calculations is given in Table-V. The signs of all the three exchange interactions obtained from the experiment and DFT calculations are found to be consistent. However, significant discrepancies are found for the strength of the exchange interactions. The experimental data reveal that the ratio $J_2/J_1$ is less than unity as compared to a high value of ~ 6 predicted by the DFT calculations. Moreover, a significant discrepancy has been found for the NNNN exchange interaction $J_3$. In the present study, the $J_3$ value is found to be similar in strength to the $J_2$ in contrast to a much weaker value (an order of magnitude smaller) predicted by DFT calculations. The simulated spin-wave spectra considering the values predicted by the DFT calculations and the magnetic structure [inplane zigzag AFM that are coupled antiferromagnetically along the $c$ axis as shown in Fig. 6(e-f)] reported in Ref. [24] is shown in Fig. 12. A stronger value of the single-ion anisotropy parameter of -0.8 meV needs to be considered to match the experimentally observed spin gap of 2 meV. The simulated dispersion modes are present between 2-7.5 meV. However, the powder averaged spectra reveal the presence of intensity up to ~ 6 meV. Although the pattern shows two excitation bands, their individual as well as overall energy range, bandwidth and intensities are significantly different from that of the experimental spectra for $Na_2Ni_2TeO_6$ [Figs. 9 and 11]. This, therefore, demands a careful DFT-based first-principles calculation, with the experimentally observed zigzag AFM structure with UUDD arrangement along the $c$ axis and a more accurate crystal structure. Moreover, proper choice of Hubbard on-site Coulomb correlations and the Hunds exchange parameter ($J_H$), and proper estimation of charge transfer energies between different orbitals, and higher plane-wave cut-off energy are required in the DFT calculations for better estimation of the strengths of the exchange interactions and understanding of microscopic magnetic properties.

The presence of gapless magnetic excitations above the $T_N$ that persist up to a high temperature is consistent with the neutron diffraction results (Fig. 6) which reveal the presence of 2D short-range spin-spin correlations up to ~ 50 K. The broad diffuse peaks in the neutron diffraction patterns appear at around $Q$ = 0.7, 1.9, and 2.6 Å$^{-1}$ [Fig. 6]. In the present INS data, gapless magnetic excitations appear over a similar $Q$ region above the $T_N$, revealing the origin as the 2D short-range magnetic ordering and spin fluctuations in the 2D magnetic ordering. It is interesting to mention that the spectral intensity at 100 K is significantly strong which is attributed to the magnetic excitation from the 2D honeycomb lattice with strong spin-spin correlations. The INS spectra, therefore, reveal that the dynamic spin correlations persist up to a high temperature, consistent with the dc susceptibility data [Fig. 4(a)] where a deviation from the paramagnetic state is evident below ~ 200 K.

**F. DISCUSSIONS:**

Now we discuss the two important findings of the present study, firstly, the coexistence of the CM and ICM AFM orderings, and secondly, the up-up-down-down (↑↑↓↓) periodicity of the CM state along the $c$ axis. The observed coexistence of the CM and ICM AFM orderings in the studied honeycomb lattice compound $Na_2Ni_2TeO_6$ is unique. Such coexisting ordering has neither been observed for any of the $Ni^{2+}$ ion ($S$=1) based related honeycomb lattice antiferromagnets $K_2Ni_2TeO_6$ [19], $Li_3Ni_2SbO_6$ [59], $Na_3Ni_2SbO_6$ [41], $Na_3Ni_2BiO_6$ and $Li_3Ni_2BiO_6$ [42], and $Cu_3Ni_2SbO_6$ [60] nor for the other related honeycomb compounds with other transition metal ions, such as, $Na_2Co_2TeO_6$ [21], $Na_3Co_2SbO_6$ [61, 62], $Li_3Co_2SbO_6$ [63, 64], $Cu_3Co_2SbO_6$ [60], and

$Ag_3Co_2SbO_6$ with $S$=3/2. For these compounds, only a single commensurate magnetic phase, viz., the 3D CM zigzag AFM ordering was reported below the respective $T_N$. On the other hand, an ICM AFM ordering was reported for the honeycomb antiferromagnet $NaNi_2BiO_{5.66}$ [65] where the origin of the ICM is described as a result of bond-dependent Kitaev-Γ-Heisenberg exchange interactions.

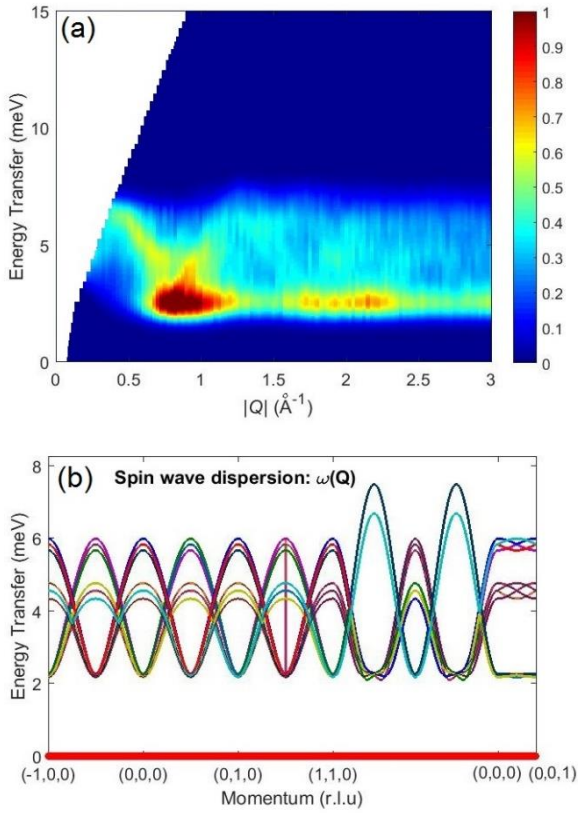

Fig. 12. (Color online) The simulated (a) powder averaged excitation spectra and (b) spin-wave dispersion modes considering the exchange constant reported by DFT calculations [24] (Table-V).

The important question is whether the coexisting of the ICM and CM orderings has resulted from the chiral symmetry of the intermediate non-magnetic Na-layer or in-plane competing $J_1$-$J_2$-$J_3$ interactions. This is to be mentioned here that the spin correlations of the ICM phase are effectively confined within the 2D honeycomb planes. The exact diagonalizations, linear spin-wave, and series expansion calculations reported that the quantum $J_1$-$J_2$-$J_3$ model on the honeycomb lattice possesses a massive degeneracy of the magnetic ground state, which might be lifted by either quantum or thermal fluctuations, the effect known as "order-by-disorder", leading to exotic ordered magnetic ground states and a complex magnetic phase diagram [9, 10, 13]. A variety of classical and quantum ground states, including the commensurate Néel, zigzag, stripy, and incommensurate spiral/helical ordered states, as well as disordered quantum spin liquid and quantum paramagnetic (plaquette valence-bond state) states, has been theoretically proposed for the $J_1$-$J_2$-$J_3$ honeycomb lattice model depending on the signs and ratios of the exchange interactions ($J_2/J_1$ and $J_3/J_1$) as well as the spin values. The CM zigzag and ICM spiral/helical phases are neighborly situated in the phase diagram. The magnetic phases are theoretically proposed to be separated by well-defined phase boundaries. However, the coexisting CM and ICM phases can occur when the system is situated close to the phase boundary as reported for the honeycomb compound γ-$BaCo_2(PO_4)_2$ [66] where the material's effective spin Hamiltonian lies near a phase boundary in the classical phase diagram and it is reported that the two magnetic orders arise likely from different spatial regions in the sample. However, for the studied compound, the derived set of the exchange constant values (Appendix-A) reveal that the effective spin Hamiltonian lies well inside the zigzag phase in the phase diagram [Fig. 13(a)] [13]. Therefore, the origin of the coexisting of the ICM and CM orderings may be ruled out due to the in-plane competing $J_1$-$J_2$-$J_3$ interactions. On the other hand, the chiral symmetry of the ICM phase is closely related to the Coulomb field that is revealed by the nuclear density distribution of the intermediate Na layer [24]. Moreover, our RMC analysis [Fig. 8] indicates that the incommensurability of the ICM phase is along the $c$ axis. The incommensurate modulation of the in plane AFM spin ordering along the $c$ axis may, therefore, be due to modulation in the exchange interactions $J_4$ that occurs through the intermediated Na-layer having chiral structure and/or due to an additional 2$^{nd}$ nearest-neighbor interlayer couplings along the $c$ axis. However, the exact variation of the $J_4$ (i.e., the origin of the ICM phase) or the possible contribution of the 2$^{nd}$ nearest-neighbor interlayer couplings (likely to be very small) cannot be evaluated using the available data set. A systematic single crystal neutron scattering study is required in this regard.

Now we focus on the second important finding in $Na_2Ni_2TeO_6$ i.e., the UUDD (↑↑↓↓) spin arrangement of the observed CM zigzag AFM state along the $c$ axis. The UUDD (↑↑↓↓) spin arrangement is an unique feature of the Ni-based compounds with the space group $P6_3/mcm$ as found for the present Na-based compound $Na_2Ni_2TeO_6$ as well as for the K-based compound $K_2Ni_2TeO_6$ [67]; and not for the isoformula compound $Na_2Co_2TeO_6$ with $P6_322$ space group [21, 68]. For $Na_2Co_2TeO_6$, an AFM i.e., UDUD (↑↓↑↓) arrangement of the zigzag planes is found. For the related layered honeycomb compounds, either a FM [UUUU(↑↑↑↑)] (for $Li_3Ni_2SbO_6$ [59] and $Na_3Ni_2BiO_6$ [42] with space group $C2/m$) or an AFM [UDUD (↑↓↑↓)] (for $Cu_3Ni_2SbO_6$ and $Cu_3Ni_2SbO_6$ [60] with space group $C2/c$) coupling of the zigzag planes were reported. The observed UUDD (↑↑↓↓) spin arrangement of the present compound with the space group $P6_3/mcm$ is exceptional concerning all the equivalent magnetic honeycomb layers constituted by $NiO_6$ and $TeO_6$ (Fig. 1). The coupling between the magnetic Ni ions from two neighbouring honeycomb layers along the $c$ axis occurs through superexchange exchange interactions $J'_4$ and $J''_4$ involving the interaction pathways Ni–O–Na2–O–Ni and Ni–O–Na1–O–Ni, respectively (Fig. 7 and Table-III). The sign and strength of such superexchange interactions are decided by the bond lengths and the bond angles of the superexchange pathways as formulated by the Goodenough-Kanamori rules [57, 58]. Usually, the directions of the superexchange pathways i.e., the directions of bond lengths and bond angles in a lattice, do not have any role in the sign of the exchange constant and magnetic symmetry. For the studied compound $Na_2Ni_2TeO_6$, the superexchange interaction pathways (Ni–O–Na2–O–Ni and Ni–O–Na1–O–Ni) are identical for all the magnetic layers and contain the same bond lengths and bond angles values. However, the orientations of these superexchange interaction pathways (Ni–O–Na2–O–Ni and Ni–O–Na1–O–Ni) are opposite in two neighboralong Na-layers which is defined by the alternating chirality of the Na-ions arrangments [Fig. 13(b)]. Interestingly, the change of the sign of the magnetic moment occurs for only one type of chirality of the Na-ion layer which leads to the UUDD (↑↑↓↓) spin arrangement along the $c$ axis i.e., a double periodicity of the magnetic spin arrangement [Fig. 13(b)]. Therefore, the chiral structure of the intermediate nonmagnetic Na-layers is responsible for the magnetic symmetry.

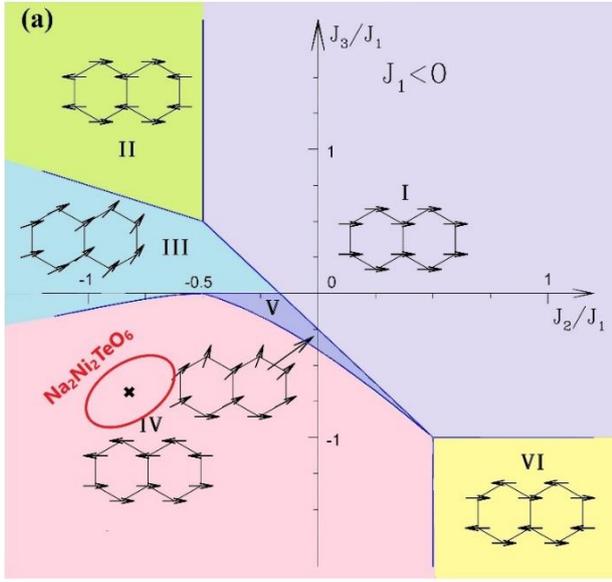

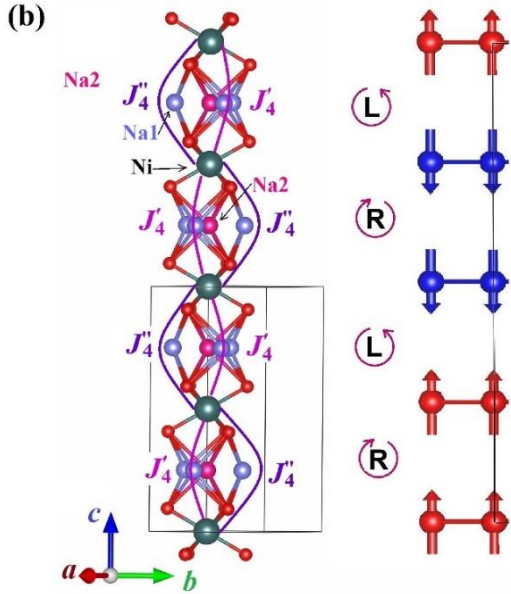

Fig. 13. (Color online) (a) The theoretical phase diagram ($J_2/J_1$ - $J_3/J_1$) for $J_1$-$J_2$-$J_3$ honeycomb lattice antiferromagnet with FM $J_1 < 0$ (adapted from Ref. [13]) with collinear and noncolinear ordered magnetic states (shown by the regions with different colors). The experimentally found zigzag ordered phase is shown by labeled IV (red region). The red ellipsoid represents the possible position of the studied compound $Na_2Ni_2TeO_6$ in the phase diagram. The "×" symbol marks the parameters used for the $S(Q,\omega)$ simulation in Fig. 11. . (b) Local magnetic coupling along the $c$ axis. The change of the orientations of the interlayer superexchange interactions pathways are evident. Corresponding change in the spin arrangement is also shown in the right side.

## IV. SUMMARY AND CONCLUSIONS:

In summary, detailed crystal structural and magnetic properties of the 2D layered spin-1 honeycomb lattice compound $Na_2Ni_2TeO_6$ have been investigated by x-ray and neutron diffraction, dc-magnetization, and inelastic neutron scattering. The layered crystal structure of $Na_2Ni_2TeO_6$ composed of magnetic layers is formed by edge shared $NiO_6$ and $TeO_6$ octahedra within the crystallographic $ab$ planes, which are well separated (~ 5.6 Å) by an intermediate Na layer along the $c$ axis. Within the magnetic layers, the honeycomb lattices are formed with spin-1 $Ni^{2+}$ ions and nonmagnetic $Te^{6+}$ ion being at the center of the honeycomb lattice. Our comprehensive study reveals a novel magnetic phenomenon where the magnetic symmetry is dictated by the intermediate nonmagnetic Na-layer, having a chiral nuclear density distribution of Na ions which is a unique feature among the Na based layered compounds, especially $A_2M_2XO_6$ or $A_3M_2XO_6$ compounds. Such chiral nuclear density distributions alternates along the $c$ axis and dictates the magnetic periodicity which results in an up-up-down-down (↑↑↓↓) spin arrangement of the inplane CM zigzag AFM structure along the $c$ axis [characterized by the propagation vector $\mathbf{k}$ = (½ ½ ½)]. Further, the CM zigzag AFM order state is found to coexist with a 2D ICM AFM state below the $T_N$ = 27.5 K. The 2D nature of the ICM AFM state is established by the RMC analyses. Above the $T_N$ = 27.5 K, a 2D ICM short-range AFM ordering is found to be present up to ~ 50 K. The spin Hamiltonian of $Na_2Ni_2TeO_6$ has been determined by an inelastic neutron scattering study and linear spin-wave analysis. The INS spectra reveal a predominant contribution from the CM zigzag AFM state. The magnetic Hamiltonian determined by the spin-wave fitting of the inelastic spectra, for the CM zigzag AFM state, reveals inplane competing exchange interactions up to 3$^{rd}$ nearest neighbors with a weak interplanar coupling and a weak single-ion anisotropy. Our results reveal that the present compound lies well inside the zigzag phase (spans over wide ranges of $J_2/J_1$ and $J_3/J_1$ values) in the theoretically proposed $J_2/J_1$ - $J_3/J_1$ phase diagram. . The present study provides a detailed microscopic understanding of coexisting CM and ICM magnetic states and divulges a novel magnetic phenomenon where the magnetic symmetry is controlled by the nonmagnetic layer.

**APPENDIX-A:**
**Spin wave analysis, extracting model solutions, and estimating uncertainty:**

We used the SpinW package to calculate the powder inelastic neutron scattering cross-section for a given Hamiltonian parameter set, based on the linear spin-wave theory (LST). The Hamiltonian involves up to five independent parameters: $J_1$, $J_2$, $J_3$, $J_4$, and $D$. In order to extracting model solution and estimating their uncertainty, simultaneous fittings of the three curves (two momentum cuts and the energy cut) are performed [Fig. 11 (e-g)] and the uncertainties are estimated by the following equation

$$\chi^2_{INS} = \sum_{i=0}^{N-1} \frac{\{I_{exp}(Q,w) - I_{cal}(Q,w)\}^2}{\sigma_i^2}$$

where, $I_{exp}(Q,w)$ and $I_{cal}(Q,w)$ are, respectively, the experimentally measured and spin-wave calculated intensities, $N$ is the number of data points, and $\sigma_i$ is the error of the $i^{th}$ experimental data point. In each step, the values $\chi^2_{INS}$ were calculated from the fittings of the three curves (two momentum cuts and the energy cut) as shown in Fig. 11 (e-g). The distribution of best fitting parameters is shown in the contour plots in Fig. 14 where the colors represent the reduced chi-square $\tilde{\chi}^2_{INS} = 1 - \chi^2_{INS}/(\chi^2_{INS})_{max}$. The optimized regions of the parameter space for the solution are shown by the ellipses. The ellipses for the optimized regions of the parameter space are determined for the solutions for which the $\tilde{\chi}^2_{INS}$ value is approximately higher than 0.8. It may be noted that some of these contours look non-ellipsoidal, which means perhaps an ellipsoid description is approximately correct (and helpful for plotting purposes). A most possible solution (marked by crosses) inside the optimized region is represented by the following parameters: $J_1$ = -1.40 meV, $J_2$ = 1.10 meV, $J_3$ = 1.00 meV, $J_4$ = 0.08 meV, and $D$ = -0.20 meV. For this solution, the $\tilde{\chi}^2_{INS}$ value is found to be 0.91.

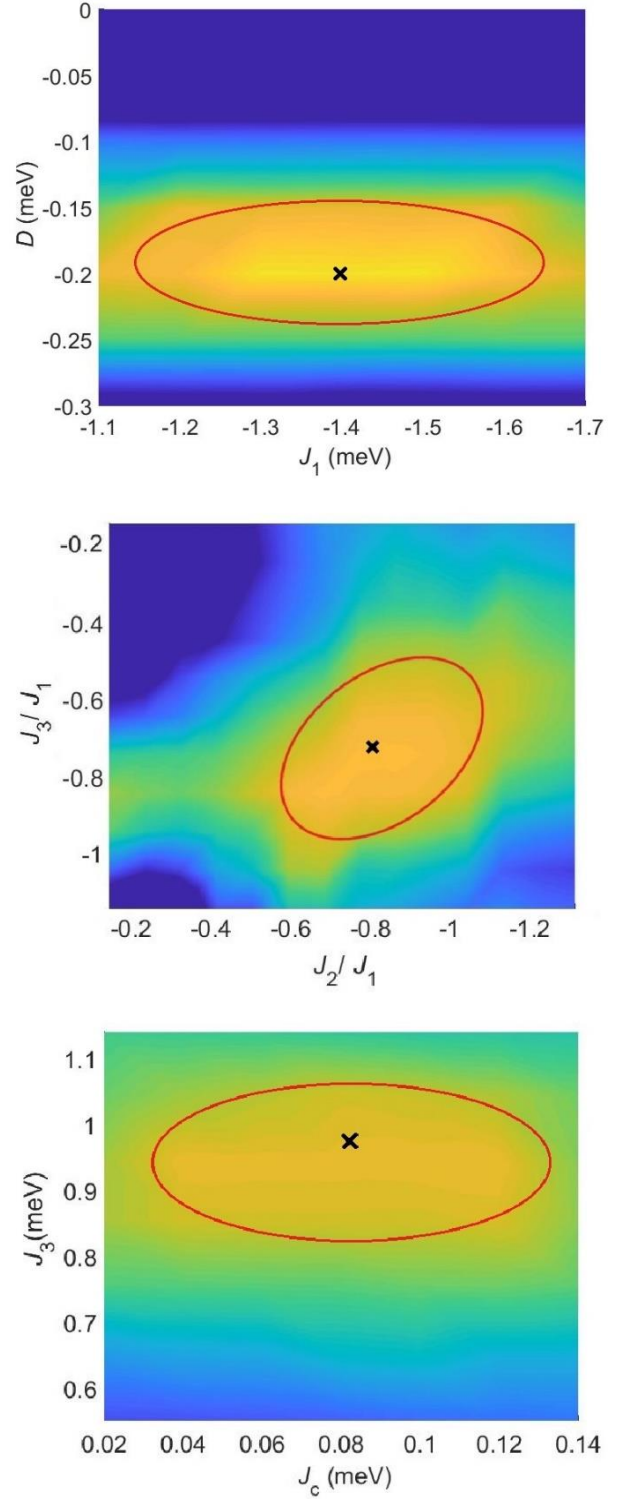

Fig. 14: Contour plots of projected $\tilde{\chi}^2_{INS}$ for Na$_2$Ni$_2$TeO$_6$ spin-wave spectrum using the parameter space of J and D values for the Hamiltonian model. The optimized regions of the parameter space for the solution are shown by the ellipses. The "×" symbol marks the parameters used for the $S(Q,\omega)$ simulation in Fig. 11. As described in the text, this model describes the main features of the magnetic excitations.

Our spin-wave analyses reveal that the observed patterns correspond to the four non-degenerate spin-wave dispersion modes with bandwidths of about 10 meV. Two distinguished dispersion bands over 2-4.5 meV and 2-12 meV are evident. Spinwave simulations reveal that the observed inplane zigzag AFM structure cannot be reproduced by a single exchange interaction; either by NN $J_1$ or by NNN $J_2$. The width of the lower energy band is dependent on the relative strength $J_2/J_1$. However, the experimental bandwidth cannot be reproduced by $J_2$ alone, and therefore, unambiguously reveal the presence of 3rd nearest-neighbor exchange interaction $J_3$. The relative widths of the energy bands over 2-5 and 2-12 meV depend on the relative strengths of the intraplanar interactions ($J_1$, $J_2$, and $J_3$). The fitting suggests that the $J_1$ is ferromagnetic (FM) and all other inplane interactions ($J_2$ and $J_3$) are antiferromagnetic. The derived values of exchange constants are consistent with the zigzag phase in the phase diagram of the $J_1$-$J_2$-$J_3$ Heisenberg honeycomb model with FM nearest-neighbor exchange interaction $J_1$ [13]. The spin-wave simulations also reveal that the AFM interplanar exchange interaction $J_4$ not only results in a dispersion along the $c$ axis but also removes the degeneracy of the dispersion modes at the bottom as well as top edges of the bands. However, the value of $J_4$ is about an order of magnitude smaller than that of the in-plane exchange interactions.

**APPENDIX-B:**
**$Na_2Ni_2TeO_6$ and Kitaev spin model**:
Kitaev spin model has recently been realized in Honeycomb based spin systems. The model features bond-dependent Ising interactions (Kitaev interactions) between spins on a honeycomb lattice. The spin-orbit coupling and electron correlations are essential for bond-dependent anisotropic interactions. Although the model was originally proposed for a highly anisotropic spin-1/2 degrees of freedom, recently, the possibility of the Kitaev spin model for the spin-1 degrees of freedom in layered transition metal oxides [$A_3Ni_2XO_6$ ($A$ = Li, Na, $X$ = Bi, Sb)] was proposed by Stavropoulos *et al.* [52]. The Kitaev interactions in spin-1 system occurs through a complex mechanism where a strong spin-orbit coupling in anion sites, which is one important ingredient for the Kitaev interaction, is expected to occur via proximity to the heavy $X$ atoms. One of the characteristic features of the 2D Kitaev model in the powder-averaged magnetic excitation pattern (including the magnetic form factor) was reported to be a non-dispersing high-energy band centred at an energy that corresponds approximately to the Kitaev exchange scale [53]. The intensity of this band is strongest at $Q=0$, and decreases with increasing $Q$. The possibility of the Kitaev interactions in the real materials such as α-RuCl$_3$ [53] and the related compounds $Na_2Co_2TeO_6$ and $Na_3Co_2SbO_6$ [54, 55] were characterized by such high energy broad mode. Especially, for these Kitaev candidate materials with the Kiaev-Heisenberg Hamiltonian, the strongest intensity of the higher energy mode is found at lower or at the same $Q$ value where low-energy spin-wave mode shows maximum intensity and the dispersion minima corresponding to the AFM zone centre. In contrast, for the studied compound $Na_2Ni_2TeO_6$, a completely different $Q$ dependence of the high energy mode (at energy transfer $\Delta E \sim$ 12 meV) is found. Here, the intensity of the high energy mode is non monotonous and strongly $Q$ dependent. The observable intensity of this mode appears only above $\sim |Q|$= 1.0 Å$^{-1}$ and becomes strongest at a higher $|Q|$ value of $\sim$ 1.5 Å$^{-1}$ which is certainly much higher than the AFM zone centre at $|Q| \sim$ 0.75 Å$^{-1}$. The absence of intensity of the high energy mode is evident at the AFM zone centre at $|Q| \sim$ 0.75 Å$^{-1}$. Further, the results reported by M. Songvilay *et al.* [54], in agreement with our spinwave simulations, reveal that the shifted (higher) $Q$ value for the intensity of the higher energy mode is a clear indication of the $J_1$-$J_2$-$J_3$ Heisenberg model. Therefore, from the $Q$ dependence of the high energy mode $\sim$ 12 meV it may be concluded that the studied compound $Na_2Ni_2TeO_6$ is better represented by the $J_1$-$J_2$-$J_3$ Heisenberg model and the additional intensity observed at $\Delta E \sim$ 12 meV and $|Q| \sim$ 1.5 Å$^{-1}$ may not be due to the anisotropic Kitaev interactions. The absence of intensity of the higher energy mode at low-$Q$ values ($|Q|<$ 1.0 Å$^{-1}$ ) further indicates the possible absence of the anisotropic Kitaev interactions in $Na_2Ni_2TeO_6$. However, the possibility of a weak Kitaev exchange interactions cannot be ruled out. In this regard, a comprehensive analysis of the INS spectrum considering a model with combined Heisenberg and Kitaev interactions is necessary to estimate the limit of the Kitaev exchange interaction. Such a study is definitely of future interest.